\documentclass[12pt, oneside]{article}
\usepackage{graphicx}
\pdfoutput=1
\usepackage[export]{adjustbox}
\usepackage[paper=a4paper, hmargin=2.6cm, vmargin=2.6cm]{geometry}
\usepackage[T1]{fontenc}
\usepackage[utf8]{inputenc}
\usepackage[lighttt]{lmodern}
\usepackage{setspace}
\usepackage[hang]{footmisc}
\usepackage{titling}
\usepackage{tocloft}
\usepackage{parskip}
\usepackage{hyperref}
\usepackage{fourier-orns}
\usepackage{booktabs}
\usepackage{shuffle}
\usepackage{amssymb}
\usepackage{tabularx}
\newcolumntype{Y}{>{\hspace{0pt}\raggedright\arraybackslash}X}
\usepackage{amsfonts}
\usepackage{amsmath}
\usepackage{amssymb}
\usepackage{mathrsfs, dsfont, amsthm}
\usepackage{mathtools, nccmath}
\usepackage{bm}
\usepackage{physics}
\allowdisplaybreaks
\numberwithin{equation}{section}

\usepackage{float}

\usepackage{booktabs, multirow}

\usepackage[dvipsnames]{xcolor}
\definecolor{dark-red}{rgb}{0.50,0.12,0.12}
\definecolor{mblue}{rgb}{0.13, 0.25, 0.55}
\definecolor{mred}{rgb}{0.70, 0.20, 0.20}
\definecolor{mgray}{rgb}{0.63, 0.63, 0.63}
\definecolor{myWhite}{RGB}{255,255,243}

\newcommand*\justify{%
  \fontdimen2\font=0.4em%
  \fontdimen3\font=0.2em%
  \fontdimen4\font=0.1em%
  \fontdimen7\font=0.1em%
  \hyphenchar\font=`\-%
}
\renewcommand{\texttt}[1]{%
  \begingroup
  \ttfamily
\begingroup\lccode`~=`/\lowercase{\endgroup\def~}{/\discretionary{}{}{}}%
\begingroup\lccode`~=`[\lowercase{\endgroup\def~}{[\discretionary{}{}{}}%
\begingroup\lccode`~=`.\lowercase{\endgroup\def~}{.\discretionary{}{}{}}%
\catcode`/=\active\catcode`[=\active\catcode`.=\active
  \justify\scantokens{#1\noexpand}%
  \endgroup
}

\usepackage{hyperref}
\hypersetup{
colorlinks=true,
allcolors=dark-red,
linktoc=page,
pdftitle={HyperPrecision},
pdfauthor={Sumit Banik, Souvik Bera},
pdfdisplaydoctitle=true,
pdfstartview=FitH
}

\usepackage[backend=biber,
            natbib=true,
            style=numeric-comp,
            maxbibnames=9,
            sorting=none,
            url=false,
            date=year,
            doi=false,
            giveninits=true,
            isbn=false,
            bibwarn=true]{biblatex}
\usepackage{orcidlink}
\definecolor{orcidlogocol}{HTML}{A6CE39}

\usepackage{tikz}
\usetikzlibrary{arrows.meta,positioning,fit,calc,patterns,decorations,decorations.markings,shapes,intersections,quotes,angles}
\usepackage[subrefformat=parens]{subcaption}

\def \d   {\mathrm{d}}

\newcommand{\ic}{\mathrm{i}}

\renewcommand{\le}{\leqslant}

\renewcommand{\ge}{\geqslant}
\renewcommand{\geq}{\geqslant}
\newcommand{\cmd}[1]{\texttt{\textcolor{mblue}{#1}}}
\usepackage{enumitem}
\setlist[itemize]{leftmargin=1.2em}

\newcommand{\eeqref}[1]{Eq.~\eqref{#1}}

\usepackage{tcolorbox}
\usepackage{colortbl}
\usepackage{listings}
\definecolor{identifiercolor}{rgb}{.4,.6,.56}
\definecolor{stringcolor}{gray}{0.5}
\definecolor{inactivecolor}{rgb}{0.2,0.2,0.2}

\begin{document}
\raggedbottom

\begin{titlingpage}
    \hfill \texttt{SLAC-PUB-260526}

    \vspace*{1em}
    \onehalfspacing
    \begin{center}
        \textbf{\Huge\texttt{HyperPrecision}}
        \\[0.5em]
        {\large A Mathematica package for High-Precision Numerical Evaluation of Multivariate Hypergeometric Functions}
        \\[0.8em]
    \end{center}
    \singlespacing
    \vspace*{1.5em}
    \begin{center}
        \textbf{Sumit Banik,\textsuperscript{1 \orcidlink{0000-0002-5869-5293}} \quad Souvik Bera\textsuperscript{2 \orcidlink{0000-0002-1784-7051}}}
    \end{center}
    \vspace*{0.5em}
    \begin{center}
        \textsl{
        $^{1}$\ SLAC National Accelerator Laboratory, Stanford University,
        \\ Stanford, California 94039, USA \\[0.5\baselineskip]
        }
        \textsl{
        $^{2}$\ Asia Pacific Center for Theoretical Physics,
        \\ Pohang, 37673, Korea \\[0.5\baselineskip]
        }
        \href{mailto:banik@stanford.edu}{\small\texttt{banik@stanford.edu}}, \quad
        \href{mailto:souvik.bera@apctp.org}{\small\texttt{souvik.bera@apctp.org}}
    \end{center}
    \vspace*{1.5em}
\begin{abstract}

\noindent

In this paper, we present \texttt{HyperPrecision}, a \textsc{Mathematica} package for high-precision numerical evaluation of general Horn-type multivariate hypergeometric functions and their Laurent expansions in a small parameter $\epsilon$. Such functions appear widely in physics and mathematics, with applications ranging from quantum field theory and string theory to number theory and statistics. Their high-precision numerical evaluation, however, remains challenging, since their defining series converge only in restricted domains and analytic continuation beyond these domains is, in general, non-trivial. \texttt{HyperPrecision} addresses this problem by automatically constructing the Pfaffian system of partial differential equations for a given hypergeometric function and restricting it to a one-dimensional contour in the space of variables connecting the starting point to the target point. The resulting ordinary differential equation is then solved by the Frobenius method, with the boundary conditions analytically determined by the defining series. We illustrate the use of the package by evaluating commonly occurring multivariate hypergeometric functions, including the Appell $F_1$, $F_2$, $F_3$, and $F_4$ functions, the Horn $G$- and $H$-series, and the Lauricella $F_A$, $F_B$, $F_C$, and $F_D$ functions, as well as by considering applications to angular integrals, Feynman integrals, and cosmological and holographic correlators.

\end{abstract}
    \vfill
\end{titlingpage}

\newpage

\section*{Program summary}
\begin{itemize}
    \item[$\diamond$] \textit{Program Title:} \texttt{HyperPrecision.wl}, version 1.2
    \item[$\diamond$] \textit{Developer's repository link:} \href{https://github.com/HyperPrecision/HyperPrecision}{\texttt{github.com/HyperPrecision/HyperPrecision}}
    \item[$\diamond$] \textit{Licensing provisions:} GNU General Public License v3
    \item[$\diamond$] \textit{Programming language:} \textsc{Wolfram Mathematica}, version 12.3 or higher
    \item[$\diamond$] \textit{External routines/libraries used:} \texttt{FiniteFlow}, \texttt{DESolver}
    \item[$\diamond$] \textit{Nature of problem:} Numerical evaluation of multivariate hypergeometric functions is challenging, since their defining series converge only in limited regions of argument space, and analytic continuation beyond these regions is generally not available in closed form for arbitrary Horn-type functions.
    \item[$\diamond$] \textit{Solution method:} Given a Horn-type hypergeometric function, the package automatically constructs the associated Pfaffian system of partial differential equations from the series definition. This system is then restricted to a one-dimensional path, reducing the problem to an ordinary differential equation. The latter is then solved numerically using the Frobenius generalised power-series method, with boundary conditions determined directly from the original hypergeometric series at the origin.
    \item[$\diamond$] \textit{Restrictions:} The current version of the package is restricted to complete Horn-type hypergeometric functions with real arguments, and its performance depends on the rank of the underlying holonomic system, the complexity of the associated Pfaffian system and on the requested numerical precision.

    \item[$\diamond$] \textit{References:}
    \begin{enumerate}
        \item[$\circ$] \href{http://www.wolfram.com/mathematica/}{\textsc{Wolfram Mathematica}}; proprietary software.
        \item[$\circ$] \href{https://github.com/peraro/finiteflow/tree/master}{\texttt{FiniteFlow}}, open-source software.
        \item[$\circ$] \href{https://gitlab.com/multiloop-pku/amflow}{\texttt{DESolver}}, open-source software.
    \end{enumerate}
\end{itemize}

\newpage
\tableofcontents
\raggedbottom
\setcounter{page}{2}


\section{Introduction}

Multivariate hypergeometric functions (MHFs)~\cite{Slater:1966,bateman,Exton:1976,Srivastava:1985,bailey1935generalized,schlosser2013multiple,aomoto2011theory,Yoshida:1997wm} are multivariable generalisations of the classical Gauss hypergeometric function ${}_2F_1$~\cite{Gauss:1813},
\begin{align}\label{def:2f1}
{}_2F_1(a,b;c;z) = \sum_{n=0}^{\infty} \frac{(a)_n (b)_n}{(c)_n} \frac{z^n}{n!} , \qquad (a)_n = \frac{\Gamma(a+n)}{\Gamma(a)} .
\end{align}
MHFs are commonly represented as multidimensional infinite series whose coefficients are products and ratios of Pochhammer symbols. A systematic study and classification of these functions can be found in~\cite{Srivastava:1985}. Some well-known classes of MHFs include the Appell~\cite{Appell:1926}, Horn~\cite{Horn:1931}, Kamp\'e de F\'eriet~\cite{KampedeFeriet:1937}, and Lauricella--Saran functions~\cite{Lauricella:1893,Saran_transhypgeo}.

\hspace{1cm} MHFs appear in a wide range of problems in mathematics and physics. In mathematics, they are central to the theory of Gel'fand--Kapranov--Zelevinsky systems of partial differential equations~\cite{Gelfand198714,Gelfand1994DiscriminantsRA,Gelfand1991HYPERGEOMETRICFT,Im1990GeneralizedEI,delaCruz:2019skx,Klausen:2019hrg,Ananthanarayan:2022ntm,Chestnov:2022alh,Feng:2019bdx}, the theory of twisted de Rham cohomology~\cite{aomoto2011theory,Yoshida:1997wm}, and the study of Mellin--Barnes (MB) integrals~\cite{Friot:2011ic,Ananthanarayan:2020fhl}. Many functions of recent interest, such as multiple polylogarithms~\cite{Goncharov:1998kja}, are also closely related to MHFs~\cite{Banik:2024ann}. In physics, MHFs arise naturally in perturbative quantum field theory, where they provide solutions to multiloop multiscale Feynman integrals~\cite{Smirnov:2012gma,Weinzierl:2002hv} that appear in scattering amplitudes. They also occur in the evaluation of cosmological correlators~\cite{Sleight:2019mgd,Chen:2024glu}, holographic correlators~\cite{Bissi:2022mrs,Bobev:2025idz} and angular integrals~\cite{Somogyi:2011ir}, among other applications.

\hspace{1cm} Since MHFs are usually defined by infinite series, these representations are valid only in restricted regions of argument space where the series converge\footnote{Even within the region of convergence, the series can converge very slowly near the boundary of the convergence region.}. Evaluating the functions beyond these regions, therefore, requires analytic continuation. For general MHFs, this is a highly non-trivial task, although several methods and software tools exist for special cases. These include approaches based on analytic continuations of the Gauss hypergeometric function ${}_2F_1$~\cite{Olsson:1964,Ananthanarayan:2021yar}, as well as methods based on MB representations~\cite{Friot:2011ic,Ananthanarayan:2020fhl,Banik:2023rrz}. However, these tools do not, in general, provide analytic continuations covering the full argument space.
For example, the recent conic-hull~\cite{Ananthanarayan:2020fhl} and triangulation~\cite{Banik:2023rrz} approaches express an MB integral as a set of series solutions that are analytic continuations of one another. For many MB integrals, however, the combined convergence regions of these series solutions do not cover the entire argument space. This leaves a region, commonly known as a \emph{white zone}, where none of the series solutions converges. Consequently, MB representations do not generally provide analytic continuations of MHFs to arbitrary target points.

\hspace{1cm} Furthermore, in many of the physics applications mentioned above, one is interested not only in the value of an MHF, but also in its Laurent expansion in a small parameter $\epsilon$, on which the Pochhammer parameters may depend. Such expansions have been extensively studied in the literature. Series expansions of one-variable hypergeometric functions are well understood~\cite{Kalmykov:2006pu,Kalmykov:2007pf,Kalmykov:2008ge,Greynat:2013hox}, while the $\epsilon$-expansion of two-variable Appell and Kamp\'e de F\'eriet functions has been discussed in~\cite{Greynat:2014jsa,Moch:2001zr,Weinzierl:halfint}. Certain MHFs arising in the computation of Feynman integrals have also been expanded using the differential-equation method~\cite{Yost:2011wk,Bytev:2012ud,Kalmykov:2020cqz}. In addition, expansions of multiple integrals and multiple sums involving hyperexponential and hypergeometric functions have been considered in~\cite{Bierenbaum:2007zu,Blumlein:2010zv,Blumlein:2011kef,Ablinger:2012ph,Schneider:2013zna,Blumlein:2021hbq}.

\hspace{1cm} Several automated packages are available for determining the expansion coefficients of certain MHFs, either analytically~\cite{Huber:2005yg,Huber:2007dx,Moch:2005uc,Weinzierl:2002hv,Ablinger:2013cf,Bera:2022ecn,Bera:2023pyz} or numerically~\cite{Huang:2012qz,Bezuglov:2025xol,Bezuglov:2025msm}. However, the scope of these packages is typically restricted to specific classes of hypergeometric functions involving only a small number of variables. As a result, a general framework for the high-precision numerical evaluation of MHFs, both directly and as a Laurent expansion in a small parameter, is still lacking.

\hspace{1cm} In this paper, we employ a strategy, inspired by the differential-equation method for Feynman integrals~\cite{Kotikov:1990kg,Gehrmann:1999as,Henn:2013pwa}, to numerically evaluate Horn-type MHFs using their systems of partial differential equations. The key observation is that Horn-type MHFs generally satisfy Pfaffian systems of differential equations of finite holonomic rank. Once such a system is known, it can be restricted to a one-dimensional contour in argument space, yielding an ordinary differential equation. Analytic continuation to the target point is then obtained by applying the Frobenius method along this contour, starting from an initial point (where the solution is known), as implemented in several numerical solvers~\cite{Hidding:2020ytt,Liu:2022chg,Armadillo:2022ugh,Prisco:2025wqs}. This idea was recently implemented for MHFs in the \texttt{PrecisionLauricella} package~\cite{Bezuglov:2025msm,Bezuglov:2025xol}, which applies the method to the numerical evaluation of the Lauricella functions $F_A^{(n)}$, $F_B^{(n)}$, and $F_D^{(n)}$ for $n \le 3$.
 In this package, however, the Pfaffian systems are not derived dynamically, but are hard-coded from~\cite{Bezuglov:2023owj}. Its applicability is therefore restricted to a small set of pre-configured MHFs, and extending it to other MHFs would require constructing and implementing the corresponding Pfaffian systems by hand. Moreover, the current version of the package cannot evaluate MHFs at points lying on singular curves.

\hspace{1cm} Many MHFs that arise in concrete physics applications lie beyond the pre-configured functions of \texttt{PrecisionLauricella}. For example, the Appell $F_4$ function (which is $F_C^{(2)}$) appears in one-loop bubble Feynman integrals~\cite{Bezuglov:2023owj}, in three-point inflationary correlators with massive exchanges~\cite{Aoki:2024uyi}, and in three-point holographic correlators in non-conformal D$p$-brane backgrounds~\cite{Bobev:2025idz}. Higher-dimensional Lauricella functions also appear in multivariate angular integrals~\cite{Lyubovitskij:2021ges,Ahmed:2024pxr} and multi-loop Feynman integrals~\cite{Berends:1993ee}. Motivated by these and other applications in physics and mathematics, we present \texttt{HyperPrecision}, a public \textsc{Mathematica}~\cite{Mathematica} package that automatically derives the Pfaffian system associated with an input MHF whose Pochhammer parameters depend linearly on $\epsilon$, and solves it numerically at arbitrary target points, including, in most cases, points lying on singular curves. The package can therefore be applied, in principle, to any Horn-type hypergeometric function with finite holonomic rank.

\hspace{1cm} The technical core of the package consists of two steps. First, the package constructs a basis for the quotient of the rational Weyl algebra by the left ideal generated by the annihilating operators of the input MHF, viewed as a finite-dimensional vector space over the field of rational functions, and computes the connection matrices on this basis using \texttt{FiniteFlow}~\cite{Peraro:2016wsq,Peraro:2019svx}. Second, the resulting Pfaffian system is restricted to a straight line in argument space connecting the starting point, taken here to be the origin, to the user-specified target point. The boundary condition at the origin is fixed analytically from the leading term of the defining series, so that no external numerical input is required. The resulting one-dimensional ordinary differential equation is then integrated to the target point using the \texttt{DESolver} package, which is a part of the \texttt{AMFlow} package~\cite{Liu:2022chg}. To obtain the $\epsilon$-expansion, the package evaluates the system on a finite grid of $\epsilon$ values and reconstructs the Laurent coefficients by interpolation, making the most expensive part of the calculation naturally parallelisable.

\hspace{1cm} The remainder of the paper is organised as follows. In Section~\ref{sec:hypgeofn}, we give a brief introduction to MHFs and the structure of their associated Pfaffian systems. In Section~\ref{sec:numerical}, we describe the automated derivation of the Pfaffian system using \texttt{FiniteFlow}, its transport along a one-dimensional contour to the target point via \texttt{DESolver}, and the reconstruction of the $\epsilon$-expansion. In Section~\ref{sec:manual}, we present the \texttt{HyperPrecision} package, including installation instructions and the main user-accessible functions, along with a worked example. In Section~\ref{sec:validation}, we validate the package against representative Feynman-integral examples, including bubble and sunset topologies. In Section~\ref{sec:Applications}, we provide three applications of the package, namely angular integrals, cosmological correlators and holographic correlators. We conclude in Section~\ref{sec:Conclusion} with a discussion of possible extensions and applications. Appendix~\ref{app:functions} documents additional package commands useful for further study, Appendix~\ref{app:seed_ints} collects expressions relevant to our cosmological correlator study, and Appendix~\ref{app:predefined} lists the predefined MHFs implemented in the package.

\section{Multivariate Hypergeometric Functions}\label{sec:hypgeofn}

This section contains a brief, self-contained introduction to the class of MHFs handled by \texttt{HyperPrecision} and fixes the notation used throughout the paper. A more comprehensive and systematic treatment of MHFs can be found in the book~\cite{Srivastava:1985}.

\subsection{Definition and Notation}
\label{subsec:hypgeo-definition}

We consider MHFs of $n$ complex variables
$\mathbf{x}=(x_1,\ldots,x_n)$ of the form
\begin{align}
    \label{eqn:mhfdefinition}
    F(\mathbf{a};\mathbf{b};\mathbf{x})
    &=
    \sum_{\mathbf{m}\in\mathbb{N}_0^n}
    \frac{(\mathbf{a})_{\mu\cdot\mathbf{m}}}
         {(\mathbf{b})_{\nu\cdot\mathbf{m}}}
    \frac{\mathbf{x}^{\mathbf{m}}}{\mathbf{m}!}
    =
    \sum_{\mathbf{m}\in\mathbb{N}_0^n}
    A(\mathbf{m})\,\mathbf{x}^{\mathbf{m}} ,
\end{align}
where $\mathbf{a}\in\mathbb{C}^p$ and $\mathbf{b}\in\mathbb{C}^q$ are vectors of Pochhammer parameters, while
$\mu\in\mathbb{Z}^{p\times n}$ and $\nu\in\mathbb{Z}^{q\times n}$ are usually integer matrices. Here $\mathbb{N}_0$ denotes the set of non-negative integers, and we use the standard multi-index notation
\begin{align}
    \mathbf{x}^{\mathbf{m}}
    :=
    \prod_{i=1}^{n} x_i^{m_i},
    \qquad
    \mathbf{m}!
    :=
    \prod_{i=1}^{n} m_i!,
    \qquad
    \sum_{\mathbf{m}\in\mathbb{N}_0^n}
    :=
    \sum_{m_1=0}^{\infty}\cdots\sum_{m_n=0}^{\infty},
\end{align}
The Pochhammer symbols are defined by
\begin{align}
    (\mathbf{a})_{\mu\cdot\mathbf{m}}
    :=
    \prod_{i=1}^{p}
    (a_i)_{(\mu\cdot\mathbf{m})_i}
    =
    \prod_{i=1}^{p}
    \frac{
      \Gamma\!\left(a_i+(\mu\cdot\mathbf{m})_i\right)
    }{
      \Gamma(a_i)
    },
\end{align}
with an analogous definition for $(\mathbf{b})_{\nu\cdot\mathbf{m}}$. The region of convergence of the series~\eeqref{eqn:mhfdefinition} is determined by Horn's theorem~\cite{Srivastava:1985}. 

\hspace{1cm} The Horn structure of MHFs is encoded in the ratios of neighbouring coefficients,
\begin{align}
    \label{eqn:Pratio}
    P_i(\mathbf{m})
    :=
    \frac{A(\mathbf{m}+\mathbf{e}_i)}{A(\mathbf{m})}
    =
    \frac{g_i(\mathbf{m})}{h_i(\mathbf{m})},
    \qquad
    i=1,\ldots,n,
\end{align}
where $\mathbf{e}_i$ is the $i$-th standard basis vector and $g_i$, $h_i$ are polynomials in the summation indices. An MHF is called \textit{Horn-type}, if $P_i(\mathbf{m})$, $i=1,\dots,n$ are rational functions of $\mathbf{m}$ and the function is called \emph{complete}~\cite{Horn:1931} if the limits
\begin{align}
    \label{eqn:completeness}
    \Phi_i(\mathbf{m})
    :=
    \lim_{t\to\infty} P_i(t\mathbf{m}),
    \qquad
    i=1,\ldots,n,
\end{align}
exist and are neither identically zero nor infinite. \texttt{HyperPrecision} works only for complete Horn-type MHFs, which include the Appell, Kampé de Fériet, and Lauricella families, as well as most MHFs that arise in Feynman integral calculus.

\subsection{Pfaffian System}\label{subsec:hypgeo-pfaffian}

It is a classical result that the MHF, as defined in~\eeqref{eqn:mhfdefinition}, satisfies a holonomic system of partial differential equations~\cite{Bateman:1953,yoshida2013fuchsian}. From the shift ratios~\eeqref{eqn:Pratio}, the function~\eeqref{eqn:mhfdefinition} is annihilated by the operators
\begin{align}\label{eqn:annihilator}
    L_i \;=\; h_i(\boldsymbol{\theta})\,\frac{1}{x_i}
            \,-\, g_i(\boldsymbol{\theta}) \, ,
    \hspace{1cm} i = 1,\dots,n \, ,
\end{align}
where $\boldsymbol{\theta} = (\theta_1,\dots,\theta_n)$, with $\theta_i = x_i \partial_{x_i}$ denoting the $i$-th Euler operator.

\hspace{1cm} For numerical calculations, it is convenient to recast the system $\{L_i \mathbf{F} = 0\}_{i=1,\dots,n}$ in Pfaffian form,
\begin{align}\label{eqn:Pfaffian}
    \d\mathbf{g} \;=\; \Omega\,\mathbf{g} \, ,
    \hspace{1cm}
    \Omega \;=\; \sum_{i=1}^{n} \Omega_i\,\d x_i \, ,
\end{align}
where the vector $\mathbf{g}$ collects $\mathbf{F}$ and a finite number of its derivatives,
\begin{align}\label{eqn:basis}
    \mathbf{g}
    \;=\; \big( \mathbf{F}\,,\; \theta_i \cdot \mathbf{F}\,,\;
                \theta_i \theta_j \cdot \mathbf{F}\,,\;
                \ldots \big)^{T} \, ,
\end{align}
and the connection matrices $\Omega_i$ are rational functions of $\mathbf{x}$ and of the parameters $\mathbf{a}$, $\mathbf{b}$. The length of $\mathbf{g}$ equals the holonomic rank $r$ of the system. The compatibility of the partial differential equations in~\eeqref{eqn:Pfaffian} is guaranteed by the integrability condition
\begin{align}\label{eqn:integrability}
    \d\Omega + \Omega \wedge \Omega \;=\; 0 \, ,
\end{align}
which is automatically satisfied by Pfaffian systems arising from holonomic MHFs. Explicit Pfaffian systems for Appell and Lauricella functions have been constructed and studied extensively in the mathematical literature~\cite{yoshida2013fuchsian,Kato_Pfaff_F4,Matsumoto_Pfaff_F4,Matsumoto_Pfaff_FA,Matsumoto_Pfaff_FD}. The algorithmic construction of~\eeqref{eqn:Pfaffian} for a general MHF, as implemented in \texttt{HyperPrecision}, is discussed in Section~\ref{sec:numerical}.

\subsection{Example: Appell \texorpdfstring{$F_2$}{F2}}
\label{subsec:hypgeo-F2}

We illustrate the definitions of Secs.~\ref{subsec:hypgeo-definition} and~\ref{subsec:hypgeo-pfaffian} with the bivariate Appell $F_2$ function
\begin{align}
    F_2(a,b_1,b_2;c_1,c_2;x,y)
    &=
    \sum_{m,n=0}^{\infty}
    \frac{(a)_{m+n}(b_1)_m(b_2)_n}
         {(c_1)_m(c_2)_n}
    \frac{x^m y^n}{m!\,n!}
    =
    \sum_{m,n=0}^{\infty} A(m,n)\,x^m y^n ,
\end{align}
which converges for $|x|+|y|<1$. In the notation introduced above, we have
$\mathbf{a}=(a,b_1,b_2)$, $\mathbf{b}=(c_1,c_2)$, and
\begin{align}
    \mu =
    \begin{pmatrix}
        1 & 1 \\
        1 & 0 \\
        0 & 1
    \end{pmatrix},
    \qquad
    \nu =
    \begin{pmatrix}
        1 & 0 \\
        0 & 1
    \end{pmatrix}.
\end{align}

The ratios of neighbouring coefficients are
\begin{align}
    P_1(m,n)
    =
    \frac{(b_1+m)(a+m+n)}
         {(m+1)(c_1+m)},
    \qquad
    P_2(m,n)
    =
    \frac{(b_2+n)(a+m+n)}
         {(n+1)(c_2+n)}.
\end{align}
Their large-order limits are
\begin{align}
    \Phi_1(m,n) = \frac{m+n}{m},
    \qquad
    \Phi_2(m,n) = \frac{m+n}{n},
\end{align}
showing that $F_2$ is complete in the sense of \eeqref{eqn:completeness}. After substituting the ratios into \eeqref{eqn:annihilator}, we get
\begin{align}
    L_1
    &=
    -a b_1
    + \big(c_1 - x(a+b_1+1)\big)\partial_x
    - b_1 y\,\partial_y
    - x y\,\partial_x\partial_y
    - (x-1)x\,\partial_{xx},
    \label{eqn:F2annh1}
    \\[2pt]
    L_2
    &=
    -a b_2
    + \big(c_2 - y(a+b_2+1)\big)\partial_y
    - b_2 x\,\partial_x
    - x y\,\partial_x\partial_y
    - (y-1)y\,\partial_{yy}.
    \label{eqn:F2annh2}
\end{align}
For this system, the holonomic rank is known to be four. We therefore choose a convenient basis of a rank-four holonomic system as
\begin{align}
    \label{eqn:F2basis}
    \mathbf{g}
    =
    \big(
        F_2,\,
        \theta_x F_2,\,
        \theta_y F_2,\,
        \theta_x\theta_y F_2
    \big)^T ,
\end{align}
where $\theta_x=x\partial_x$ and $\theta_y=y\partial_y$. In this basis, the system generated by
Eqs.~\eqref{eqn:F2annh1} and~\eqref{eqn:F2annh2} can be written in Pfaffian form as
\begin{align}
    \d\mathbf{g}
    =
    \left(\Omega_1\,\d x+\Omega_2\,\d y\right)\mathbf{g}.
\end{align}
The connection matrices are

\begin{align}
    \Omega_1 &= \left(
\begin{array}{cccc}
 0 & \frac{1}{x} & 0 & 0 \\
 -\frac{a b_1}{x-1} & \frac{a x+b_1 x-c_1+1}{x(1-x)} & -\frac{b_1}{x-1} & \frac{1}{1-x} \\
 0 & 0 & 0 & \frac{1}{x} \\
 \frac{a b_1 b_2 y}{(x-1)(x+y-1)} & \frac{b_2 y(a+b_1-c_1+1)}{(x-1)(x+y-1)} & \frac{b_1\big(b_2 y - (x-1)(a-c_2+1)\big)}{(x-1)(x+y-1)} & \frac{b_2 x y - (x-1)\big(a x + b_1 x - c_2 x + c_1 y - c_1 + x - y + 1\big)}{(x-1)\,x\,(x+y-1)} \\
\end{array}
\right), \label{eqn:F2OmegaX}
\end{align}
\begin{align}
    \Omega_2 &= \left(
\begin{array}{cccc}
 0 & 0 & \frac{1}{y} & 0 \\
 0 & 0 & 0 & \frac{1}{y} \\
 -\frac{a b_2}{y-1} & -\frac{b_2}{y-1} & \frac{a y+b_2 y-c_2+1}{y(1-y)} & \frac{1}{1-y} \\
 \frac{a b_1 b_2 x}{(y-1)(x+y-1)} & \frac{b_2\big(b_1 x - (y-1)(a-c_1+1)\big)}{(y-1)(x+y-1)} & \frac{b_1 x(a+b_2-c_2+1)}{(y-1)(x+y-1)} & \frac{b_1 x y - (y-1)\big(a y + b_2 y + c_2 x - c_1 y - c_2 - x + y + 1\big)}{(y-1)\,y\,(x+y-1)} \\ \label{eqn:F2OmegaY}
\end{array}
\right).
\end{align}
We will later use this Pfaffian system, together with the boundary condition determined from the leading term of the defining series at the origin, as a working example in Section~\ref{subsec:example}. In the derivative basis used here, the Pochhammer parameters appear exclusively in the numerators of the connection matrices in Eqs.~\eqref{eqn:F2OmegaX} and~\eqref{eqn:F2OmegaY}. We have observed the same pattern for all MHFs considered so far.

\section{Numerical Evaluation of Hypergeometric Functions}
\label{sec:numerical}
In this section, we provide an overview of the working principle of \texttt{HyperPrecision}, which can be divided into two parts. First, for a given MHF, the package derives the annihilating operators introduced in Section~\ref{subsec:hypgeo-pfaffian} and then constructs the Pfaffian system dynamically, using finite-field methods implemented in \texttt{FiniteFlow}~\cite{Peraro:2016wsq,Peraro:2019svx}. The resulting Pfaffian system is then restricted to a one-dimensional contour and solved numerically with \texttt{DESolver}~\cite{Liu:2022chg}, after which the Laurent expansion in the regulator $\epsilon$ is reconstructed by interpolation. We will discuss each of the above steps in more detail next.

\subsection{Deriving the Pfaffian System}
\label{subsec:derivingPfaffian}

As discussed in Section~\ref{sec:hypgeofn}, the annihilating operators of an MHF can be obtained directly from its defining series. In principle, the corresponding system of partial differential equations can be brought into Pfaffian form using Gr\"obner basis methods. This can be done, for example, with \texttt{HolonomicFunctions}~\cite{KoutschanPhD,koutschan2010fast} in \textsc{Mathematica}, or with \texttt{ConnectionMatrices}~\cite{ConnectionMatricesSource} in \textsc{Macaulay2}. There are other alternative methods to derive Pfaffian systems, notably the Macaulay-matrix method of~\cite{Chestnov:2022alh} and the $D$-module techniques of~\cite{Henn:2023tbo}. Our reduction is similar in spirit to these.

\hspace{1cm}
In \texttt{HyperPrecision}, we instead employ a more direct reduction
procedure that is well-suited for a finite-field implementation. Starting
from the annihilating operators, we generate a large system of equations by
repeated differentiation. Assuming that the input MHF has finite holonomic
rank, sufficiently many such equations allow the higher derivatives to be
expressed in terms of a finite set of lower derivatives. These lower
derivatives can then be chosen as basis elements, whose number equals the
holonomic rank of the system. Once the basis has been identified, the
derivatives of the basis elements with respect to the variables can likewise
be reduced to the basis itself. The resulting coefficients define the entries
of the Pfaffian connection matrices.

\hspace{1cm} For non-trivial multivariate functions, this reduction problem can involve
large linear systems, and a purely symbolic computation may suffer from severe
intermediate expression swell. We therefore carry out the reduction with
\texttt{FiniteFlow}~\cite{Peraro:2016wsq,Peraro:2019svx}, using modular
arithmetic and rational reconstruction to obtain the exact rational connection
matrices. We observed that, within this approach, the computation time to obtain the Pfaffian system is significantly reduced compared to conventional Gr\"obner basis methods. To quantify this improvement, Table~\ref{tab:time_pfaffian} compares the times required to compute the Pfaffian systems for the Lauricella functions $F_D^{(n)}$, with $n=1,\dots,5$, using conventional Gr\"obner basis methods and our \texttt{FiniteFlow} based approach.

\begin{table}[h]
  \centering
  \footnotesize
  \setlength{\tabcolsep}{5pt}
  \begin{tabular}{lcccc}
    \toprule
    Function & Rank & \texttt{HyperPrecision} & \texttt{HolonomicFunctions} & \texttt{ConnectionMatrices} (\texttt{M2}) \\
    \midrule
    Gauss $\,_2F_1$        & 2 & $0.11$ s & $0.0012$ s & $0.015$ s \\
    Appell $F_1$           & 3 & $0.15$ s            & $0.036$ s            & $0.2$ s \\
    Lauricella $F_D^{(3)}$ & 4 & $0.3$ s            & $0.6$ s           & $52$ s \\
    Lauricella $F_D^{(4)}$ & 5 & $1.4$ s            & $6$ s & $> 24 $ hr  \\
    Lauricella $F_D^{(5)}$ & 6 & $25$   s          & $73$ s  & $-$  \\
    \bottomrule
  \end{tabular}
  \caption{Computation time of the connection matrices of Lauricella $F_D^{(n)}$. As  there is no command in the \texttt{HolonomicFunctions} package that finds the Pfaffian matrices directly, the numbers under the \texttt{HolonomicFunctions} column refer to the computation time of the Gr\"obner basis calculations using the \texttt{OreGroebnerBasis} command, which is the most time-consuming step of the Pfaffian matrix calculation.}
  \label{tab:time_pfaffian}
\end{table}

\subsection{Numerical Evaluation Using the Frobenius Method}
\label{subsec:transport}

Once the Pfaffian system has been constructed, the problem is reduced to transporting its solution from a point where the boundary value is known to the target point
\(\mathbf{x}_0=(x_{0,1},\ldots,x_{0,n})\). We choose the origin as the boundary point, since it lies in the convergence region of the defining series. At this point, the leading term of Eq.~\eqref{eqn:mhfdefinition} gives \(F(\mathbf{0})=1\), while all other basis elements generated by Euler operators \(\theta_{x_i}=x_i\partial_{x_i}\) vanish. Thus, the basis vector satisfies
\begin{align}
    \label{eqn:boundarycondition}
    \mathbf{g}(\mathbf{0})
    =
    (1,0,0,\ldots,0)^T .
\end{align}

Next, to perform the transport, we reduce the multivariate Pfaffian system to an ordinary differential equation by introducing an auxiliary variable \(\eta\),
\begin{align}
    \label{eqn:etaparam}
    x_i(\eta)
    =
    \frac{x_{0,i}}{1+\eta},
    \qquad
    i=1,\ldots,n .
\end{align}
With this parametrisation, the limit \(\eta\to\infty\) corresponds to the origin (starting point), while \(\eta=0\) gives the target point \(\mathbf{x}_0\). The path is therefore a straight contour in \(\mathbf{x}\)-space from the boundary point to the point of evaluation. Therefore, restricting the Pfaffian system~\eeqref{eqn:Pfaffian} to the contour of~\eeqref{eqn:etaparam} gives the one-dimensional system
\begin{align}
    \label{eqn:etaODE}
    \frac{\d\mathbf{g}}{\d\eta}
    =
    M(\eta)\,\mathbf{g},
    \qquad
    M(\eta)
    =
    \sum_{i=1}^{n}
    \frac{\partial x_i}{\partial\eta}
    \Omega_i\big(\mathbf{x}(\eta)\big)
    =
    -\sum_{i=1}^{n}
    \frac{x_{0,i}}{(1+\eta)^2}
    \Omega_i\big(\mathbf{x}(\eta)\big).
\end{align}
This construction is analogous in spirit to the auxiliary-mass-flow method of~\cite{Liu:2017jxz,Liu:2022chg} where the variable \(\eta\) acts as the flow parameter, the boundary condition is imposed at \(\eta=\infty\), and the desired value is obtained by transporting the solution to \(\eta=0\). In practice, when only one target point is required, \texttt{HyperPrecision} constructs the restricted matrix \(M(\eta)\) directly along this contour without first forming the full multivariate Pfaffian system. This default behaviour is described in Appendix~\ref{app:FindRestrictedPfaffianSystem}.

\hspace{1cm} Finally, the ordinary differential equation in \eeqref{eqn:etaODE} is solved with the 
package \texttt{DESolver}~\cite{Liu:2022chg} shipped with \texttt{AMFlow}. 
Given \(M(\eta)\) and the boundary condition 
in~\eeqref{eqn:boundarycondition}, \texttt{DESolver} constructs Frobenius 
generalised power-series ans\"atze around the relevant singular points 
and matches neighbouring expansions in their common region of 
convergence. This is the standard strategy used in numerical 
generalised-series methods~\cite{Moriello:2019yhu,Lee:2017qql,Lee:2018ojn,Hidding:2020ytt}. 
The matching points and truncation orders are chosen internally to reach 
the working precision requested by the user. The result of the transport 
is the numerical vector \(\mathbf{g}(\eta=0)\), whose first component is 
the value of the hypergeometric function at \(\mathbf{x}=\mathbf{x}_0\).

\hspace{1cm} The procedure described so far gives the value of the MHF for a fixed numerical value of the regulator \(\epsilon\). To obtain the Laurent expansion around \(\epsilon=0\), \texttt{HyperPrecision} avoids a symbolic expansion of the Pfaffian system. Instead, following~\cite{Liu:2022mfb,Liu:2022chg,Bezuglov:2025msm,Bezuglov:2025xol}, it evaluates the function at a finite set of numerical \(\epsilon\) values and reconstructs the Laurent coefficients by interpolation. This reconstruction is performed internally using the \texttt{EpsFit[]} command of \texttt{DESolver}, which chooses the sampling points according to the requested order in \(\epsilon\) and the target precision, calls the transport once for each value of \(\epsilon\), and reconstructs the coefficients of the Laurent expansion. Since the evaluations at different \(\epsilon\) values are independent, this stage is naturally parallelisable.

\hspace{1cm} Therefore, for a given MHF, once the Pfaffian system has 
been constructed using the method of 
Section~\ref{subsec:derivingPfaffian}, \texttt{HyperPrecision} can compute 
the Laurent expansion at the target point \(\mathbf{x}_0\) automatically. We next describe how to set up 
\texttt{HyperPrecision} and illustrate its use on a simple example in the following section.

\section{Manual}
\label{sec:manual}

The aim of this section is to provide the minimal information needed to set up \texttt{HyperPrecision}, load its dependencies, and run the main user-accessible routines.

\subsection{Installation}
\label{subsec:installation}

\texttt{HyperPrecision} is publicly available at
\begin{center}
\href{https://github.com/HyperPrecision/HyperPrecision}{\texttt{github.com/HyperPrecision/HyperPrecision}}
\end{center}
and is distributed under the GNU General Public License v3. It is written in \textsc{Mathematica} and has been tested with version 12.3 and above.

\hspace{1cm} \texttt{HyperPrecision} relies on two external libraries, 
\texttt{FiniteFlow}~\cite{Peraro:2016wsq,Peraro:2019svx} and 
\texttt{DESolver}~\cite{Liu:2022chg}, each corresponding to one of the 
two main steps of the algorithm described in Section~\ref{sec:numerical}. 
The version of \texttt{DESolver} required by \texttt{HyperPrecision} 
differs from the public release in a small number of internal routines. 
For this reason, the file \texttt{DESolver\_patch} is shipped
with 
\texttt{HyperPrecision} to properly interface the latter with the original \texttt{DESolver}, which is also shipped for convenience.
\texttt{FiniteFlow}, on the other hand, must be installed by following 
the instructions in its official 
\href{https://github.com/peraro/finiteflow}{repository}. Once 
it is installed, the user can clone the 
\texttt{HyperPrecision} repository and load the package from 
\textsc{Mathematica} with
\begin{lstlisting}[extendedchars=true,language=Mathematica,literate={`}{{\textasciigrave}}1]
SetDirectory["/path/to/HyperPrecision"];
Get["HyperPrecision`"]
\end{lstlisting}
where \texttt{/path/to/HyperPrecision} should be replaced by the absolute path to the package directory. When the package is loaded, the original version of \texttt{DESolver}, the patched file \texttt{DESolver\_patch} and the installed \texttt{FiniteFlow} interface are loaded automatically. The repository also contains the notebook \texttt{Examples.nb}, which provides examples of the top-level commands described below and in the appendices.

When loaded, \texttt{HyperPrecision} also launches a pool of parallel sub-kernels to distribute evaluations across the $\epsilon$-grid (see the \cmd{ParallelRun} option). By default, the number of sub-kernels launched is twice the number of available physical processor cores. This number can be overridden by assigning a value to the global variable \texttt{\$KernelLaunch} before loading the package with \texttt{Get[]}; for example, \texttt{\$KernelLaunch = 8;}. If a parallel pool is already running, no additional sub-kernels are launched.

\subsection{Top-Level Command: \cmd{HypExpand[]}}\label{subsec:hypexpand}

The first and main top-level command of \texttt{HyperPrecision.wl} is \cmd{HypExpand[]}. It takes as input a Horn-type hypergeometric series, a kinematic point, and a target precision, and returns the Laurent expansion of the series in $\epsilon$, evaluated at the target point.

\begin{lstlisting}[language=Mathematica]
HypExpand[HypSeries, ParSub, VarSub, {Eps, Ord}, NPrec]
\end{lstlisting}

\paragraph{Description:} Computes the Laurent expansion in \texttt{Eps} up to order \texttt{Ord} of the hypergeometric series \texttt{HypSeries}, evaluated at the target point specified by \texttt{VarSub} to \texttt{NPrec} correct digits.

\paragraph{Arguments:} \texttt{HypSeries} is the hypergeometric series given in terms of Pochhammer parameters, with summation indices detected automatically. \texttt{ParSub} is the list of substitution rules for the Pochhammer parameters. \texttt{VarSub} is the list of rules \texttt{Var $\to$ TargetPoint} specifying the kinematic variables and the evaluation point. \texttt{Eps} is the symbol of the Laurent expansion parameter, and \texttt{Ord} is the highest order of the Laurent expansion to be computed. \texttt{NPrec} sets the target number of correct digits in the final numerical result.

\paragraph{Return value:} The list of Laurent coefficients from the lowest non-vanishing order to \texttt{Ord}.

\paragraph{Options:}
\cmd{HypExpand[]} accepts the following options:
\begin{itemize}
    \item \cmd{IDelta} fixes the side of the branch cut on which the target point is evaluated when the straight-line transport contour from the initial point to the target crosses a real singularity of the Pfaffian system. The accepted values are $\pm I$, with the default value $-I$. The two prescriptions select different branches of the function and, in general, lead to different numerical results.

    \item \cmd{VerboseMode} controls the amount of progress information printed during the computation. The default value is \texttt{False}. When set to \texttt{True}, the package prints diagnostic messages and integration progress.

    \item \cmd{ParallelRun} controls whether the evaluations on the $\epsilon$-grid are distributed across parallel sub-kernels. The default value is \texttt{True}. Setting it to \texttt{False} forces the computation to run serially on the main kernel.

    \item \cmd{PreRestriction} controls how the differential equation is constructed. The default value is \texttt{True}: the system is first restricted to the one-dimensional contour, and only the univariate matrix $M(\eta)$ is derived via \cmd{FindRestrictedPfaffianSystem[]}. This approach is efficient for a single target point and remains feasible at high holonomic rank. Setting the option to \texttt{False} causes the package to derive the full multivariate Pfaffian system via \cmd{FindPfaffianSystem[]} before transport. Because this system is independent of the target point, it can be reused for multiple target points of the same function.

    \item \cmd{SummedBoundary} controls how the boundary condition for the transport is obtained. When the option is set to its default value, \texttt{False}, the boundary condition is determined from the behaviour exactly at the origin. When it is set to \texttt{True}, the boundary condition is instead obtained by summing the defining series at a point very close to the origin. Both procedures yield the same result.

    \item \cmd{WPrecision} sets the internal working precision. When the option is set to its default value, \texttt{Automatic}, the working precision is inferred from \texttt{NPrec}. Setting it to a positive integer specifies the working precision explicitly.
\end{itemize}

\paragraph{Internals:} \cmd{HypExpand[]} is a top-level wrapper. With the default setting \cmd{PreRestriction}~$\to$~\texttt{True}, it calls \cmd{FindRestrictedPfaffianSystem[]} (Appendix~\ref{app:FindRestrictedPfaffianSystem}) to construct the restricted contour matrix $M(\eta)$ directly. It then calls \cmd{TransportDE[]} (Appendix~\ref{app:TransportDE}) to transport the solution from the origin to the target point and reconstruct the $\epsilon$-expansion. With \cmd{PreRestriction}~$\to$~\texttt{False}, it first calls \cmd{FindPfaffianSystem[]} (Appendix~\ref{app:FindPfaffianSystem}) to derive the full multivariate Pfaffian system and then passes that system to \cmd{TransportDE[]}.

\subsection{Top-Level Command: \cmd{HypFunctionExpand[]}}\label{subsec:hypfunctionexpand}

The second top-level command is \cmd{HypFunctionExpand[]}. It provides a convenient interface for the most frequently encountered MHFs, sparing the user the need to write the defining series in terms of Pochhammer symbols.

\begin{lstlisting}[language=Mathematica]
HypFunctionExpand[Func, {Eps, Ord}, NPrec]
\end{lstlisting}

\paragraph{Description:} Computes the Laurent expansion in \texttt{Eps} up to order \texttt{Ord} of the function \texttt{Func} to \texttt{NPrec} correct digits, with the Pochhammer parameters and target point supplied as arguments to \texttt{Func}.

\paragraph{Arguments:} \texttt{Func} is a call to a recognised hypergeometric function head with all parameters and kinematic variables filled in. The currently supported heads are \texttt{Hypergeometric2F1} and \texttt{HypergeometricPFQ} (the latter restricted to $q = p-1$), the Appell functions \texttt{AppellF1}--\texttt{AppellF4}, the Horn $G$-series \texttt{HornG1}--\texttt{HornG3}, the Horn $H$-series \texttt{HornH1}--\texttt{HornH7}, and the Lauricella series \texttt{LauricellaFA}--\texttt{LauricellaFD} in an arbitrary number of variables. The series definition and call signature of each head can be retrieved with the standard \textsc{Mathematica} usage query, e.g.~\texttt{?HornH4} or \texttt{?LauricellaFA}. \texttt{Eps}, \texttt{Ord}, and \texttt{NPrec} have the same meaning as in \cmd{HypExpand[]}.

\paragraph{Return value:} The list of Laurent coefficients from the lowest non-vanishing order to \texttt{Ord}.

\paragraph{Options:} \cmd{HypFunctionExpand[]} accepts the same options as \cmd{HypExpand[]}.

\paragraph{Internals:} \cmd{HypFunctionExpand[]} pattern-matches on the head of \texttt{Func}, extracts the parameters and kinematic arguments, looks up the defining series of the corresponding function from the internal table, and forwards the call to \cmd{HypExpand[]} (see Section~\ref{subsec:hypexpand}). 

\subsection{Example: Appell \texorpdfstring{$F_2$}{F2}}\label{subsec:example}

We illustrate the use of \texttt{HyperPrecision} on the Appell $F_2$ function introduced as a running example in Section~\ref{subsec:hypgeo-F2}, evaluated at the target point $(x,y) = (3,\,11/3)$ with parameters $a=2$, $b_1=3/2$, $b_2 = 1+\epsilon$, $c_1 = 4$, $c_2 = -1-\epsilon$, and with its Laurent expansion computed up to order $\epsilon^1$ with $10$ correct digits. 

\hspace{1cm} Since $(x,y) = (3,\,11/3)$ lies outside the convergence region $|x|+|y|<1$ of the defining series, the straight-line transport contour from the origin to the target crosses a real singularity of the Pfaffian system. Therefore, the complex value of the result reflects the branch selected by the default prescription \cmd{IDelta}~$\to -I$. The required parameter and variable substitution rules,
\begin{lstlisting}[language=Mathematica,literate={->}{{$\to$}}1 {\\[Epsilon]}{{$\epsilon$}}1]
F2Par = {a -> 2, b1 -> 3/2, b2 -> 1 + \[Epsilon], c -> 4, d -> -1 - \[Epsilon]};
F2Var = {x -> 3, y -> 11/3};
\end{lstlisting}
together with the symbolic series (only the summand)
\begin{lstlisting}[language=Mathematica]
F2Series = (Pochhammer[a, m + n] Pochhammer[b1, m] Pochhammer[b2, n] x^m y^n) /
           (Pochhammer[c, m] Pochhammer[d, n] m! n!);
\end{lstlisting}
are shared by all three forms of the calculation below.

\paragraph{Direct evaluation with \cmd{HypExpand[]}:}
The most direct route passes the symbolic series, the substitution rules, and the target precision to \cmd{HypExpand[]}:
\begin{lstlisting}[language=Mathematica,literate={->}{{$\to$}}1 {\\[Epsilon]}{{$\epsilon$}}1]
In[]:=  HypExpand[F2Series, F2Par, F2Var, {\[Epsilon], 1}, 10]
Out[]=  (0.528662817 - 4.194390019 I) + 0.5149686376/\[Epsilon]
        - (10.978138236 + 4.834942296 I) \[Epsilon]
\end{lstlisting}

\paragraph{Evaluation with \cmd{HypFunctionExpand[]}:}
The same evaluation can be performed without writing out the defining series, but by using the built-in command \texttt{AppellF2}:
\begin{lstlisting}[language=Mathematica,literate={->}{{$\to$}}1 {\\[Epsilon]}{{$\epsilon$}}1]
In[]:=  HypFunctionExpand[AppellF2[2, 3/2, 1 + \[Epsilon], 4, -1 - \[Epsilon], 3, 11/3], {\[Epsilon], 1}, 10]
Out[]=  (0.528662817 - 4.194390019 I) + 0.5149686376/\[Epsilon]
        - (10.978138236 + 4.834942296 I) \[Epsilon]
\end{lstlisting}

\paragraph{Two-step evaluation with \cmd{FindPfaffianSystem[]} and \cmd{TransportDE[]}.}
Finally, users who need finer control over the intermediate Pfaffian system (for instance, to inspect the connection matrices or to supply a custom basis) may explicitly call the \cmd{FindPfaffianSystem[]} and \cmd{TransportDE[]} commands, whose full usage is documented in Appendices~\ref{app:FindPfaffianSystem} and~\ref{app:TransportDE}, respectively.

The Pfaffian system associated with $F_2$ is obtained as follows,
\begin{lstlisting}[language=Mathematica,literate={->}{{$\to$}}1]
In[]:=  F2Pfaffian = FindPfaffianSystem[F2Series, {m, n}, {x, y}];
\end{lstlisting}
where the basis and connection matrices returned in \texttt{F2Pfaffian} coincide with those given in Eqs.~\eqref{eqn:F2basis}, \eqref{eqn:F2OmegaX} and~\eqref{eqn:F2OmegaY}. The system is then transported to the target point by
\begin{lstlisting}[language=Mathematica,literate={->}{{$\to$}}1 {\\[Epsilon]}{{$\epsilon$}}1]
In[]:=  TransportDE[F2Pfaffian, F2Par, F2Var, {\[Epsilon], 1}, 10]
Out[]=  (0.528662817 - 4.194390019 I) + 0.5149686376/\[Epsilon]
        - (10.978138236 + 4.834942296 I) \[Epsilon]
\end{lstlisting}
Therefore, the above three routes return identical numerical values. Further worked examples covering the Horn $G$ and $H$ series and the Lauricella functions are collected in the ancillary notebook \texttt{Examples.nb} in the \texttt{HyperPrecision} repository.

\section{Validation of the Package}\label{sec:validation}

In this section, we benchmark \texttt{HyperPrecision} against a set of representative MHFs whose numerical evaluation can be performed using publicly available software. For the Appell functions $F_1$, $F_2$, and $F_3$, we perform extensive cross-checks against the \texttt{PrecisionLauricella} package, and an in-house \texttt{C++} ordinary differential equation solver based on the Bulirsch–Stoer algorithm, implemented in \texttt{Boost} \cite{boostcpp}
for the $\epsilon$-expansion, and against the packages \texttt{AppellF1.wl}~\cite{Bera:2024hlq}, \texttt{AppellF2.wl}~\cite{Ananthanarayan:2021bqz}, and \texttt{AppellF3.wl} for the finite terms in the $\epsilon$-expansion. A representative comparison for Appell $F_2$ is discussed in Section~\ref{sec:testonF2}. 

\hspace{1cm} For hypergeometric functions in three variables, analogous tests are carried out for the Lauricella functions $F_A$, $F_B$ and $F_D$ using \texttt{PrecisionLauricella} for the $\epsilon$-expansion and the \texttt{LauricellaFD.wl}, \texttt{LauricellaSaranFS} packages~\cite{Bera:2024hlq} for the finite terms in the $\epsilon$-expansion. We further validate a selection of connection formulae for one- and two-variable hypergeometric functions, as detailed in Section~\ref{Sec:testreductionformula}. While the package is primarily designed for regular points, it can, in certain cases, produce correct results at singular points as well; this behaviour is discussed in Section~\ref{sec:onsingularity}. Finally, since no publicly available packages exist for Lauricella $F_C^{(N)}$, we validate the implementation of Lauricella $F_C^{(3)}$, along with Appell $F_4$, against banana-family Feynman integrals evaluated using \texttt{AMFlow}~\cite{Liu:2022chg}, as detailed in Section~\ref{subsec:bananaFC}.

\subsection{Test on Appell \texorpdfstring{$F_2$}{F2}}\label{sec:testonF2}

\begin{figure}
    \centering
    \includegraphics[width=1\linewidth]{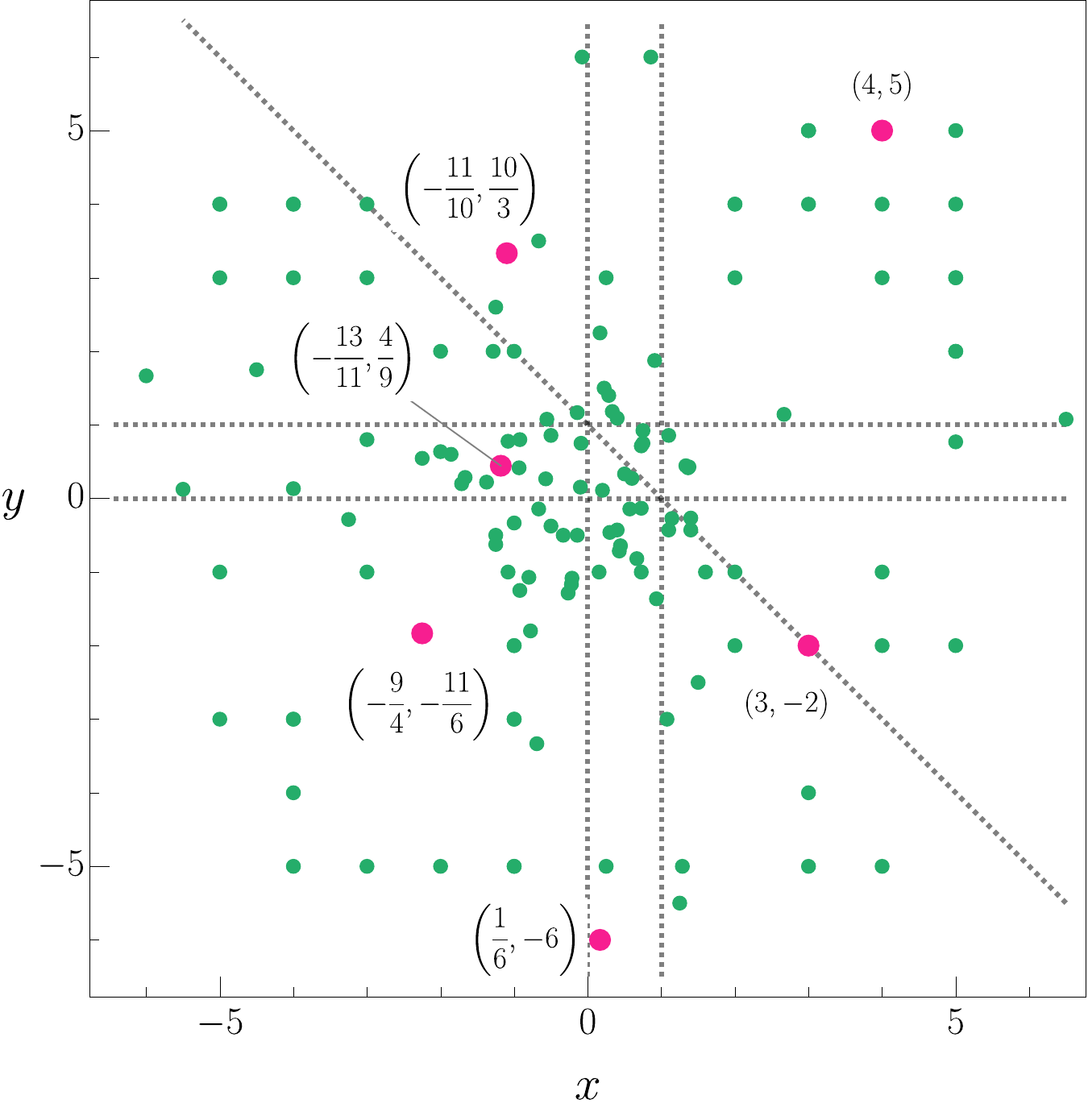}
    \caption{Random sampling of $(x, y)$ points used to validate the $\epsilon$-expansion in \eeqref{eqn:F2test} of Appell $F_2$ against \texttt{PrecisionLauricella}. The points we checked are shown as green dots, with six points highlighted in red corresponding to the points given in Table~\ref{tab:F2result}. The dashed lines represent the singular lines of $F_2$.}
    \label{fig:f2}
\end{figure}

We consider the $\epsilon$-expansion of the following Appell $F_2$ function,
\begin{align}\label{eqn:F2test}
    F_2\left(\frac{6}{5}+3\epsilon,\frac{7}{6}-\frac{2\epsilon}{3},\frac{3}{2}+\epsilon;-\frac{1}{2}+\frac{4\epsilon}{3},-\frac{2}{7}+\epsilon;x,y\right) = \sum_{n=0}^{3} c_n\, \epsilon^n,
\end{align}
evaluated at 150 randomly chosen real values of $(x, y)$. The coefficients $c_n$ are computed independently using \texttt{HyperPrecision} and \texttt{PrecisionLauricella}, with a target precision of $15$ correct digits in each case, and the resulting numerical values are compared. A representative set of six sample points, together with the corresponding coefficients, is presented in Table~\ref{tab:F2result}. These points are also shown in Figure~\ref{fig:f2}, highlighted in red. Whenever both packages return numerical results, complete agreement is observed. In a small number of cases, however, \texttt{PrecisionLauricella} fails to complete the computation and returns \texttt{\$Aborted}. The zeroth-order coefficient $c_0$ can additionally be evaluated using the \texttt{AppellF2.wl} package, and full agreement at $\mathcal{O}(\epsilon^0)$ is obtained for all nonsingular points. For points lying on the singular locus of $F_2$, such as the final example shown in the table, \texttt{AppellF2.wl} instead returns \texttt{Indeterminate}.

Analogous tests have been performed for additional parameter choices, though these are not presented explicitly. Similar comparisons have been carried out for other two- and three-variable hypergeometric functions, yielding consistent results in all cases.



\setlength{\tabcolsep}{4pt}
\renewcommand{\arraystretch}{1.25}

\begin{table}[!ht]
\centering
\tiny
\caption{%
  Numerical coefficients (15 significant digits) of the
  $\epsilon$-expansion of $F_2$ (\eeqref{eqn:F2test}).
  \texttt{\$Aborted} indicates a computation that did not complete or threw an error, and
  \texttt{Indeterminate} signals that the point lies on the singular line.
  In all cells where a comparison is possible, the \texttt{HyperPrecision} value agrees with the available reference value to all 15 digits shown.
}
\label{tab:F2result}
\begin{tabularx}{\textwidth}{c c Y Y Y}
\toprule
$(x,\,y)$ & $n$ & \texttt{AppellF2.wl} & \texttt{HyperPrecision} & \texttt{PrecisionLauricella} \\
\midrule

\multirow{4}{*}{$(4,\;5)$}
 & $0$ & $-1.62474116759515 + 0.927990517063951\,\mathrm{i}$
       & $-1.62474116759515 + 0.927990517063951\,\mathrm{i}$
       & $-1.62474116759515 + 0.927990517063951\,\mathrm{i}$ \\
 & $1$ &
       & $9.43535176688389 + 13.0560628870546\,\mathrm{i}$
       & $9.43535176688389 + 13.0560628870546\,\mathrm{i}$ \\
 & $2$ &
       & $46.1590605860182 - 46.1436055332228\,\mathrm{i}$
       & $46.1590605860182 - 46.1436055332228\,\mathrm{i}$ \\
 & $3$ &
       & $-152.009899919865 - 82.7531523975762\,\mathrm{i}$
       & $-152.009899919865 - 82.7531523975762\,\mathrm{i}$ \\
\midrule

\multirow{4}{*}{$\bigl(-\tfrac{11}{10},\;\tfrac{10}{3}\bigr)$}
 & $0$ & $-1112.54606625148 - 50.1274805523579\,\mathrm{i}$
       & $-1112.54606625148 - 50.1274805523579\,\mathrm{i}$
       & $-1112.54606625148 - 50.1274805523579\,\mathrm{i}$ \\
 & $1$ &
       & $-4337.07564527505 + 10311.8653426463\,\mathrm{i}$
       & $-4337.07564527505 + 10311.8653426463\,\mathrm{i}$ \\
 & $2$ &
       & $30522.5982910366 + 37876.4194343631\,\mathrm{i}$
       & $30522.5982910366 + 37876.4194343631\,\mathrm{i}$ \\
 & $3$ &
       & $105147.862520380 + 12106.8000698106\,\mathrm{i}$
       & $105147.862520380 + 12106.8000698106\,\mathrm{i}$ \\
\midrule

\multirow{4}{*}{$\bigl(-\tfrac{13}{11},\;\tfrac{4}{9}\bigr)$}
 & $0$ & $9.15734059903726$
       & $9.15734059903726$
       & $9.15734059903726$ \\
 & $1$ &
       & $31.5780190936921$
       & $31.5780190936921$ \\
 & $2$ &
       & $49.1347137076479$
       & $49.1347137076479$ \\
 & $3$ &
       & $-32.7856608624096$
       & $-32.7856608624096$ \\
\midrule

\multirow{4}{*}{$\bigl(-\tfrac{9}{4},\;-\tfrac{11}{6}\bigr)$}
 & $0$ & $0.792005709611611$
       & $0.792005709611611$
       & $0.792005709611611$ \\
 & $1$ &
       & $1.62627472317645$
       & $1.62627472317645$ \\
 & $2$ &
       & $7.52525345589954$
       & $7.52525345589954$ \\
 & $3$ &
       & $29.4786197938007$
       & $29.4786197938007$ \\
\midrule

\multirow{4}{*}{$\bigl(\tfrac{1}{6},\;-6\bigr)$}
 & $0$ & $0.0429676545922663$
       & $0.0429676545922663$
       & \multirow{4}{*}{\texttt{\$Aborted}} \\
 & $1$ &
       & $-0.357573631298948$
       & \\
 & $2$ &
       & $2.46290755994306$
       & \\
 & $3$ &
       & $-1.31365950314284$
       & \\
\midrule

\multirow{4}{*}{$(3,\;-2)$ } \\
\multirow{4}{*}{(lies on the singular line)} 

 & $0$ & \texttt{Indeterminate}
       & $0.649962181091589 + 0.0637137025057460\,\mathrm{i}$
       & \multirow{4}{*}{\texttt{\$Aborted}} \\
 & $1$ &
       & $2.17268990374748 - 0.755887546899809\,\mathrm{i}$
       & \\
 & $2$ &
       & $2.07677950304340 + 0.977967330594271\,\mathrm{i}$
       & \\
 & $3$ &
       & $40.5216001175546 - 13.2609710069217\,\mathrm{i}$
       & \\
\bottomrule
\end{tabularx}
\end{table}

\subsection{Test on Reduction Formulae} \label{Sec:testreductionformula}
For certain special values of the Pochhammer parameters, MHFs reduce to considerably simpler closed-form expressions, involving logarithms and polylogarithms. The following identities, taken from~\cite{Bera:2023pyz}, provide a set of such reductions for the Appell and Lauricella functions:
\begin{small}
\begin{align}
F_1(1,1,1;2;x,y) &= \frac{\ln(1-y) - \ln(1-x)}{x - y} \\[6pt]
F_2(1,1,1;2,2;x,y) &= \frac{\ln(1-x)}{y} - \frac{\ln(1-x)}{xy} - \frac{\ln\!\left(1 - \frac{x}{1-y}\right)}{x} \notag \\
&\quad - \frac{\ln\!\left(1 - \frac{x}{1-y}\right)}{y} + \frac{\ln\!\left(1 - \frac{x}{1-y}\right)}{xy} - \frac{\ln(1-y)}{y} \\[6pt]
F_3(1,1,1,1;2;x,y) &= \frac{\ln(1-x) + \ln(1-y)}{xy - x - y} \\[6pt]
F_4(1,1;1,1;x,y) &= \frac{1}{\sqrt{1 - 2x + x^2 - 2y - 2xy + y^2}} \\[6pt]
F_D(1,1,1,1;2;x,y,z) &= \frac{-x\ln(1-x)}{(x-y)(x-z)} + \frac{-y\ln(1-y)}{(x-y)(y-z)} + \frac{-z\ln(1-z)}{(x-z)(z-y)} \\[6pt]
F_S(1,1,1,1,1;2;x,y,z) &= \frac{-x\ln(1-x)}{(xy-x-y)(xz-x-z)} + \frac{-y\ln(1-y)}{(xy-x-y)(y-z)} + \frac{z\ln(1-z)}{(xz-x-z)(y-z)}
\end{align}
\end{small}
We validate \texttt{HyperPrecision} against these identities by sampling random values of $(x, y)$ and $(x, y, z)$ for the two- and three-variable cases, respectively, finding complete numerical agreement in all cases.

\subsection{Test on Connection Formulae}

\subsubsection{Euler Transformation of Gauss \texorpdfstring{$_2F_1$}{2F1}}

We validate \texttt{HyperPrecision} against well-known transformation formulae. One such example is the Euler transformation of the Gauss hypergeometric function,
\begin{align}
    {}_{2}F_{1}(a, b; c; z) = (1-z)^{-a}\, {}_{2}F_{1}\!\left(a, c-b; c; \frac{z}{z-1}\right), \label{PET1}
\end{align}
where both sides are evaluated numerically and compared. We note that, in order to obtain consistent results, the left-hand side must be evaluated with \texttt{IDelta} $\to -I$ (resp.\ $+I$) and compared against the right-hand side evaluated with \texttt{IDelta} $\to +I$ (resp.\ $-I$), since the transformation $z \to \frac{z}{z-1}$ inverts the branch cut convention.

\subsubsection{Appell \texorpdfstring{$F_3$--$F_2$}{F3-F2} Connection Formula}

We also validate the connection formula relating Appell $F_3$ to Appell $F_2$~\cite{Srivastava:1985},
\begin{align}
&F_3\left(a_1, a_2, b_1, b_2;\, c;\, x, y\right) \nonumber\\
&= f(a_1, a_2, b_1, b_2;\, x, y) \cdot F_2\!\left(a_1 + a_2 - c + 1,\, a_1,\, a_2;\; a_1 - b_1 + 1,\, a_2 - b_2 + 1;\; \tfrac{1}{x},\, \tfrac{1}{y}\right) \nonumber\\[4pt]
&+ f(a_1, b_2, b_1, a_2;\, x, y) \cdot F_2\!\left(a_1 + b_2 - c + 1,\, a_1,\, b_2;\; a_1 - b_1 + 1,\, b_2 - a_2 + 1;\; \tfrac{1}{x},\, \tfrac{1}{y}\right) \nonumber\\[4pt]
&+ f(b_1, a_2, a_1, b_2;\, x, y) \cdot F_2\!\left(a_2 + b_1 - c + 1,\, b_1,\, a_2;\; b_1 - a_1 + 1,\, a_2 - b_2 + 1;\; \tfrac{1}{x},\, \tfrac{1}{y}\right) \nonumber\\[4pt]
&+ f(b_1, b_2, a_1, a_2;\, x, y) \cdot F_2\!\left(b_1 + b_2 - c + 1,\, b_1,\, b_2;\; b_1 - a_1 + 1,\, b_2 - a_2 + 1;\; \tfrac{1}{x},\, \tfrac{1}{y}\right),
\end{align}
where
\begin{align}
    f(\lambda, \mu, \rho, \sigma;\, x, y)
= (-x)^{-\lambda}(-y)^{-\mu}\,
\frac{\Gamma(c)\,\Gamma(\rho - \lambda)\,\Gamma(\sigma - \mu)}
     {\Gamma(\rho)\,\Gamma(\sigma)\,\Gamma(c - \lambda - \mu)}.
\end{align}
As in the previous case, consistent numerical agreement requires that the left-hand side be evaluated with \texttt{IDelta} $\to -I$ (resp.\ $+I$) and compared against the right-hand side evaluated with \texttt{IDelta} $\to +I$ (resp.\ $-I$), since the transformation $z \to 1/z$ applied to both arguments $z \in \{x, y\}$ inverts the branch cut convention.

\subsubsection{Appell \texorpdfstring{$F_3$--$F_1$}{F3-F1} Relation}

We next consider the following relation between Appell $F_3$ and Appell $F_1$~\cite{Srivastava:1985},
\begin{align}
    F_3(a, a'; b, c-b; c; x, y) = (1-y)^{-a'} \cdot F_1\!\left(b;\, a, a';\, c;\, x, \frac{y}{y-1}\right),
\end{align}

in which only one of the arguments undergoes a transformation. We find that, unlike the previous cases, numerical consistency cannot be established for such formulae within the current framework. This is because correctly handling transformations acting on individual arguments, or those of a higher-order nature, requires full knowledge of the complex structure of the arguments, which lies beyond the scope of the present version of \texttt{HyperPrecision}. Extending the package to support such cases is planned for future work.
\subsection{Test on Lauricella
\texorpdfstring{$F_C^{(N)}$}{FCN}
from Two-Loop Feynman Integrals}\label{subsec:bananaFC}

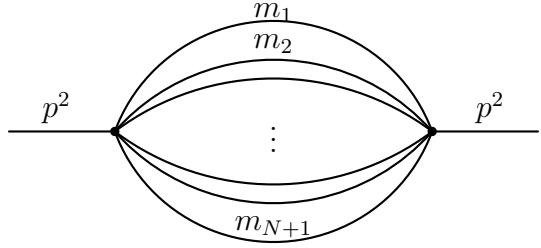
\begin{figure}[ht]
\centering

\begin{tikzpicture}[scale=0.7,thick, line cap=round, line join=round, >=Latex]

\def\a{3.0}

\newcommand{\TopArc}[3]{%
    \pgfmathsetmacro{\r}{#1}%
    \pgfmathsetmacro{\c}{sqrt(\r*\r-\a*\a)}%
    \pgfmathsetmacro{\ang}{atan(\c/\a)}%
    \draw (-\a,0) arc[start angle={180-\ang}, end angle={\ang}, radius=\r];
    \ifx&#2&%
    \else
      \node at (0,{\r-\c + #3}) {$#2$};
    \fi
}

\newcommand{\BottomArc}[3]{%
    \pgfmathsetmacro{\r}{#1}%
    \pgfmathsetmacro{\c}{sqrt(\r*\r-\a*\a)}%
    \pgfmathsetmacro{\ang}{atan(\c/\a)}%
    \draw (-\a,0) arc[start angle={180+\ang}, end angle={360-\ang}, radius=\r];
    \ifx&#2&%
    \else
      \node at (0,{-\r+\c - #3}) {$#2$};
    \fi
}

\draw (-5.0,0) -- (-\a,0);
\node at (-4.1,0.45) {$p^2$};

\draw (\a,0) -- (5.0,0);
\node at (4.1,0.45) {$p^2$};

\fill (-\a,0) circle (2.5pt);
\fill (\a,0) circle (2.5pt);

\TopArc{5.0}{m_1}{1.25}
\TopArc{4.0}{m_2}{0.30}
\TopArc{3.2}{}{0}

\BottomArc{3.2}{}{0}
\BottomArc{4.0}{}{0}
\BottomArc{5.0}{m_{N+1}}{0.80}

\node at (0,0) {$\vdots$};


\end{tikzpicture}

\caption{$N$-loop banana integral}
\label{fig:banana}
\end{figure}

In this subsection, we consider the $N$-loop banana Feynman integral (Figure~\ref{fig:banana}) in $D$ dimensions,
\begin{small}
\begin{align}
    I_{1,1,\dots,N+1}(p^2, m_1^2, \dots, m_{N+1}^2 )
    \;=\;
    \int \prod_{i=1}^N \!\left(\frac{\d^D k_i}{i \pi^{D/2}} \right)
    \prod_{i=1}^N \!\left(\frac{1}{k_i^2 - m_i^2}\right)
    \frac{1}{\big(p-\sum_{i=1}^N k_i\big)^2 - m_{N+1}^2}\,.
\end{align}
\end{small}
The solution of this integral is known in closed form in terms of Lauricella $F_C^{(N)}$~\cite{Berends:1993ee}, defined in \eeqref{eqn:LauricellaFC}:
\begin{small}

\begin{align}\label{eqn:bananaFC}
&I_{1,1,\dots,N+1}(p^2, m_1^2, \dots, m_{N+1}^2 ) = \notag \\[6pt]
&\quad=\;
(-1)^{N+1}
\left(m_{N+1}^2\right)^{\!\frac{N(D-2)-2}{2}} 
\sum_{k=0}^{N}
\frac{
  \Gamma\!\left(\dfrac{D - k(D-2)}{2}\right)
  \Gamma\!\left(\dfrac{2 - k(D-2)}{2}\right)
}{
  \Gamma\!\left(\dfrac{D}{2}\right)
}\;
\Gamma\!\left(\frac{D-2}{2}\right)^{\!k}
\Gamma\!\left(\frac{2-D}{2}\right)^{\!N-k} \notag \\[6pt]
&\qquad\times\;
\sum_{\substack{\mathcal{I}\,\subseteq\,\{1,\ldots,N\}\\|\mathcal{I}|\,=\,k}}
\left(
  \prod_{i=1}^{N} z_i^{\frac{D-2}{2}}
  \prod_{j\in\mathcal{I}} z_j^{-\frac{D-2}{2}}
\right)
F_C^{(N)}\!\left(
  \frac{D-k(D-2)}{2},\;
  \frac{2-k(D-2)}{2};\;
  \big\{c_j^{(\mathcal{I})}\big\}_{j=1}^{N+1};\;
  z_1,\ldots,z_{N+1}
\right) \,,
\end{align}
\end{small}
where the Pochhammer parameters in the denominator of $F_C^{(N)}$ are
\begin{equation}
  c_j^{(\mathcal{I})} \;=\;
  \begin{cases}
    \dfrac{4-D}{2} & \text{if } j \in \mathcal{I}, \\[10pt]
    \dfrac{D}{2}   & \text{if } j \notin \mathcal{I},
  \end{cases}
  \qquad j = 1, \ldots, N+1\,,
\end{equation}
and $z_i = m_i^2/m_{N+1}^2$ for $i = 1,\dots,N$, $z_{N+1} = p^2/m_{N+1}^2$.

\subsubsection{One-Loop Bubble and Appell \texorpdfstring{$F_4$}{F4}}
Let us consider the one-loop bubble with $m_2^2 = 1$:
\begin{align}\label{eqn:bubble}
    I^{\rm bub}_{1,1}(p^2, m_1^2 , m_2^2 = 1)
    \;=\; \int \frac{\d^D k}{i \pi^{D/2}} \frac{1}{(k^2 - m_1^2)\,((p-k)^2 - 1)}\,.
\end{align}
This integral can be expressed in terms of $F_C^{(2)}$ or the Appell $F_4$ function:
\begin{align} \label{eq:bubble_F4}
    I^{\rm bub}_{1,1}(p^2, m_1^2 , m_2^2 = 1)
    & = z_1^{1-\epsilon}\, \Gamma(\epsilon -1)\, F_4\!\left(1,2-\epsilon ;2-\epsilon ,2-\epsilon;\,z_1,z_2\right) \nonumber\\
    & \quad + \frac{\Gamma(\epsilon)\,\Gamma(1-\epsilon)}{\Gamma(2-\epsilon)}\, F_4\!\left(1,\epsilon ; \epsilon ,2-\epsilon;\,z_1,z_2\right)\,\,
\end{align}
where $z_1 = m_1^2$ and $z_2 = p^2$. 

\begin{figure}
    \centering
    \includegraphics[width=1\linewidth]{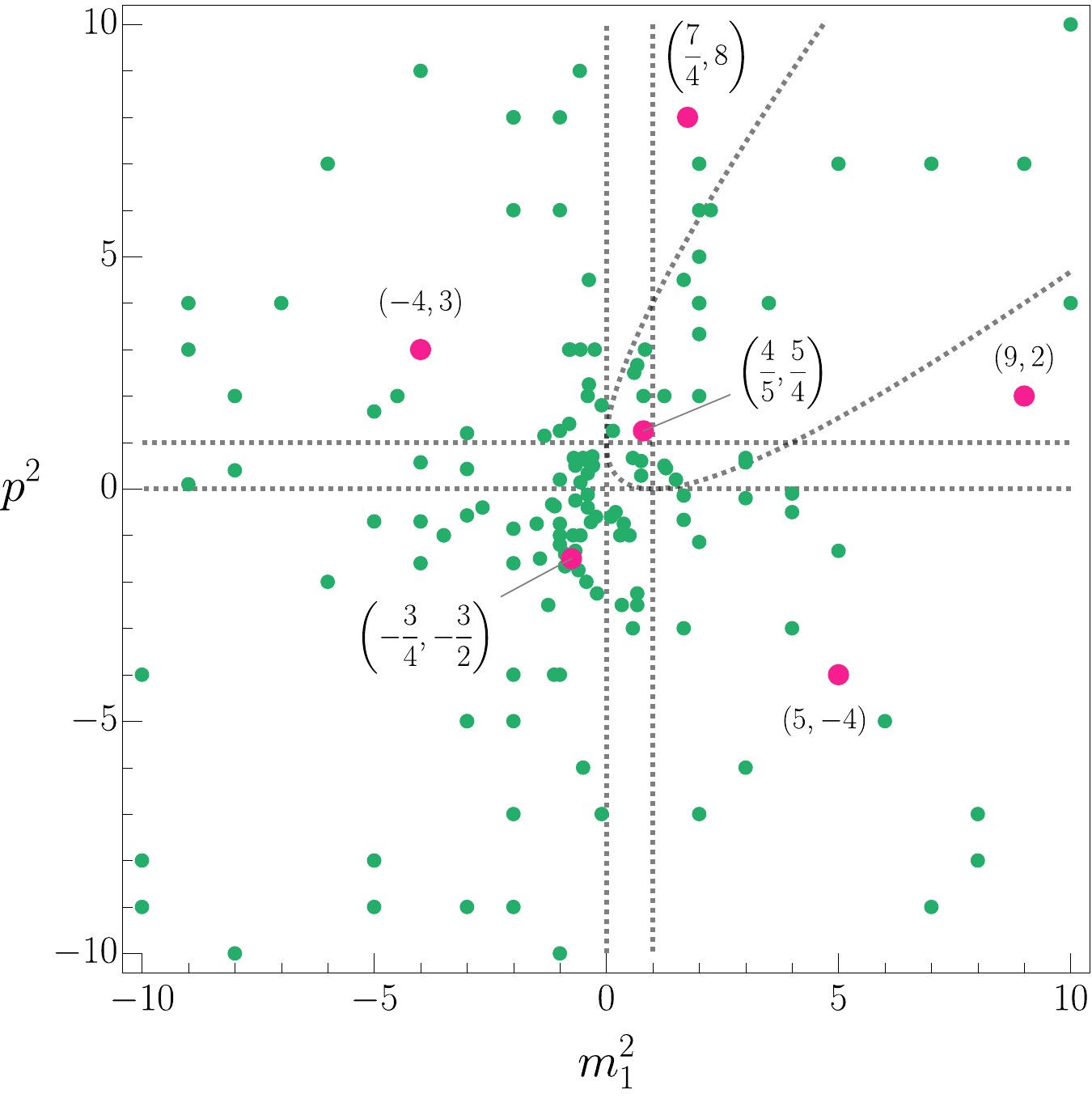}
    \caption{Random sampling of kinematic points $(m_1^2, p^2)$ used to validate the one-loop bubble integral in \eeqref{eqn:bubble} against the Appell $F_4$ representation in \eeqref{eq:bubble_F4}. The kinematic points we checked are shown as green dots, with six points highlighted in red corresponding to the points given in Table~\ref{tab:bubble_F4}. The dashed lines represent the singular lines of Appell $F_4$.}
    \label{fig:bubble_f4}
\end{figure}

\hspace{1cm} The bubble integral has been evaluated numerically using \texttt{AMFlow}~\cite{Liu:2022chg}, while the Appell $F_4$ functions appearing on the right-hand side of \eeqref{eq:bubble_F4} are computed via \texttt{HyperPrecision} with the prescription \texttt{IDelta} $\to +I$, with a target precision of $30$ correct digits. To validate the results, we randomly sample 150 kinematic points in the $(m_1^2, p^2)$ plane, indicated as dots in Figure~\ref{fig:bubble_f4}, and find perfect agreement between 
the two approaches across all sampled points. A representative subset of six such kinematic points, highlighted in red in the figure, is presented along with the corresponding numerical results in Table~\ref{tab:bubble_F4}.

{
It is useful to explain how the branch is chosen in this example, since this also illustrates the physical role of \texttt{IDelta}. The bubble integral is real below the normal threshold $p^2=(m_1+m_2)^2$ and develops an absorptive part above it, where the two internal lines can go on shell. The Feynman prescriptions $p^2\to p^2+i0$ and $m_i^2\to m_i^2-i0$ determine the sign of this imaginary part by specifying the side from which the singular locus is approached. Under the change of variables to the Appell $F_4$ arguments, $z_1=m_1^2$ and $z_2=p^2$, these prescriptions become $\operatorname{Im}z_1=-0$ and $\operatorname{Im}z_2=+0$. The singular locus is defined by the vanishing of the K\"all\'en function $\Delta=\lambda(p^2,m_1^2,m_2^2)$. At the normal threshold, $\partial_{p^2}\Delta>0$ and $\partial_{m_i^2}\Delta<0$. The two contributions therefore reinforce each other, giving $\operatorname{Im}\Delta>0$ as the transport contour crosses the singular locus. This fixes the evaluation to a unique sheet. We find that this physical branch corresponds to \texttt{IDelta} $\to +I$. With this setting, the $F_4$ representation reproduces the \texttt{AMFlow} result.
}

\begin{table}[ht!]
\centering
\tiny
\caption{%
    Numerical coefficients $c_n$ of the $\epsilon$-expansion $\left( \sum_n c_n\,\epsilon^n \right)$ of Eqs.~\eqref{eqn:bubble} and~\eqref{eq:bubble_F4}. The imaginary contributions of order $10^{-41}$, shown in red in the \texttt{HyperPrecision} column, are numerical noise and are absent in the \texttt{AMFlow} values.
}
\label{tab:bubble_F4}
\begin{tabularx}{\textwidth}{c c X X}
\toprule
\small{$(m_1^2,\,p^2)$} & \small{$n$} & \small{Numerical Value of \eeqref{eqn:bubble} using \texttt{AMFlow}} & \small{Numerical Value of the $F_4$ representation of the bubble integral in \eeqref{eq:bubble_F4} using \texttt{HyperPrecision}} \\
\midrule

\multirow{6}{*}{$(9,\;2)$}
 & $-1$             & $1$
                    & $1 - \textcolor{red}{1.3079159382974498\times10^{-41}\,\mathrm{i}}$ \\
 & $\phantom{-}0$   & $-1.9659378964334402$
                    & $-1.9659378964334402 + \textcolor{red}{3.4027776666360004\times10^{-41}\,\mathrm{i}}$ \\
 & $\phantom{-}1$   & $\phantom{-}2.9278221055325533$
                    & $\phantom{-}2.9278221055325533 - \textcolor{red}{5.631856098640424\times10^{-41}\,\mathrm{i}}$ \\
 & $\phantom{-}2$   & $-3.6048301800706084$
                    & $-3.6048301800706084 + \textcolor{red}{7.431053157823472\times10^{-41}\,\mathrm{i}}$ \\
 & $\phantom{-}3$   & $\phantom{-}4.0584743024516134$
                    & $\phantom{-}4.0584743024516134 - \textcolor{red}{8.705454736391553\times10^{-41}\,\mathrm{i}}$ \\
 & $\phantom{-}4$   & $-4.330218704953727$
                    & $-4.330218704953727 + \textcolor{red}{9.509213091589668\times10^{-41}\,\mathrm{i}}$ \\
\midrule

\multirow{6}{*}{$\bigl(\tfrac{4}{5},\;\tfrac{5}{4}\bigr)$}
 & $-1$             & $1$ & $1$ \\
 & $\phantom{-}0$   & $-0.19696764924345078$ & $-0.19696764924345078$ \\
 & $\phantom{-}1$   & $\phantom{-}0.8530298699667874$ & $\phantom{-}0.8530298699667874$ \\
 & $\phantom{-}2$   & $-0.5666001973386617$ & $-0.5666001973386617$ \\
 & $\phantom{-}3$   & $\phantom{-}0.7132865764960987$ & $\phantom{-}0.7132865764960987$ \\
 & $\phantom{-}4$   & $-0.6723564167406063$ & $-0.6723564167406063$ \\
\midrule

\multirow{6}{*}{$\bigl(\tfrac{7}{4},\;8\bigr)$}
 & $-1$             & $1$
                    & $1 - \textcolor{red}{1.5055038261413344\times10^{-41}\,\mathrm{i}}$ \\
 & $\phantom{-}0$   & $\phantom{-}0.3754221130493028 + 1.7807290488089067\,\mathrm{i}$
                    & $\phantom{-}0.3754221130493028 + 1.7807290488089067\,\mathrm{i}$ \\
 & $\phantom{-}1$   & $-1.4272757337091522 + 0.8525362269998027\,\mathrm{i}$
                    & $-1.4272757337091522 + 0.8525362269998027\,\mathrm{i}$ \\
 & $\phantom{-}2$   & $-0.9466124730646993 - 0.6282359381703576\,\mathrm{i}$
                    & $-0.9466124730646993 - 0.6282359381703576\,\mathrm{i}$ \\
 & $\phantom{-}3$   & $\phantom{-}0.09363533592176071 - 0.6118853762442324\,\mathrm{i}$
                    & $\phantom{-}0.09363533592176071 - 0.6118853762442324\,\mathrm{i}$ \\
 & $\phantom{-}4$   & $\phantom{-}0.22538508503381993 - 0.1192897787208112\,\mathrm{i}$
                    & $\phantom{-}0.22538508503381993 - 0.1192897787208112\,\mathrm{i}$ \\
\midrule

\multirow{6}{*}{$(-4,\;3)$}
 & $-1$             & $1$
                    & $1 + \textcolor{red}{2.1712927295533245\times10^{-41}\,\mathrm{i}}$ \\
 & $\phantom{-}0$   & $-0.9869709058395669 + 2.7285269151830613\,\mathrm{i}$
                    & $-0.9869709058395669 + 2.7285269151830613\,\mathrm{i}$ \\
 & $\phantom{-}1$   & $-2.405753001351794 - 3.2861998736666638\,\mathrm{i}$
                    & $-2.405753001351794 - 3.2861998736666638\,\mathrm{i}$ \\
 & $\phantom{-}2$   & $\phantom{-}3.6026340573675393 + 0.8250977832995108\,\mathrm{i}$
                    & $\phantom{-}3.6026340573675393 + 0.8250977832995108\,\mathrm{i}$ \\
 & $\phantom{-}3$   & $-2.5745835756959954 + 0.2777608210909248\,\mathrm{i}$
                    & $-2.5745835756959954 + 0.2777608210909248\,\mathrm{i}$ \\
 & $\phantom{-}4$   & $\phantom{-}2.041876842688367 - 0.09769242266166027\,\mathrm{i}$
                    & $\phantom{-}2.041876842688367 - 0.09769242266166027\,\mathrm{i}$ \\
\midrule

\multirow{6}{*}{$\bigl(-\tfrac{3}{4},\;-\tfrac{3}{2}\bigr)$}
 & $-1$             & $1$
                    & $1 + \textcolor{red}{3.501269109115439\times10^{-41}\,\mathrm{i}}$ \\
 & $\phantom{-}0$   & $\phantom{-}0.27568887655228397 + 0.8249670983578458\,\mathrm{i}$
                    & $\phantom{-}0.27568887655228397 + 0.8249670983578458\,\mathrm{i}$ \\
 & $\phantom{-}1$   & $\phantom{-}0.025907709262957525 + 0.6458099301784577\,\mathrm{i}$
                    & $\phantom{-}0.025907709262957525 + 0.6458099301784577\,\mathrm{i}$ \\
 & $\phantom{-}2$   & $-0.7382126356863302 + 0.01835083332026289\,\mathrm{i}$
                    & $-0.7382126356863302 + 0.01835083332026289\,\mathrm{i}$ \\
 & $\phantom{-}3$   & $\phantom{-}0.27147548950952033 - 0.1570759390760943\,\mathrm{i}$
                    & $\phantom{-}0.27147548950952033 - 0.1570759390760943\,\mathrm{i}$ \\
 & $\phantom{-}4$   & $-0.33046690376761934 - 0.06562619360896026\,\mathrm{i}$
                    & $-0.33046690376761934 - 0.06562619360896026\,\mathrm{i}$ \\
\midrule

\multirow{6}{*}{$(5,\;{-4})$}
 & $-1$             & $1$
                    & $1 + \textcolor{red}{2.2360679774997897\times10^{-41}\,\mathrm{i}}$ \\
 & $\phantom{-}0$   & $-1.805282722131563$ & $-1.805282722131563$ \\
 & $\phantom{-}1$   & $\phantom{-}2.535879173489941$ & $\phantom{-}2.535879173489941$ \\
 & $\phantom{-}2$   & $-3.0040179600724946$ & $-3.0040179600724946$ \\
 & $\phantom{-}3$   & $\phantom{-}3.300322866917357$ & $\phantom{-}3.300322866917357$ \\
 & $\phantom{-}4$   & $-3.4691241311893823$ & $-3.4691241311893823$ \\
\bottomrule
\end{tabularx}
\end{table}

\subsubsection{Two-Loop Sunset and Lauricella \texorpdfstring{$F_C^{(3)}$}{FC3}}
For the two-loop sunset integral
\begin{align}\label{eqn:sunset}
    I^{\rm sun}_{1,1,1}(p^2, m_1^2 , m_2^2, m_3^2 = 1)
    \;=\; \int \frac{\d^D k_1}{i \pi^{D/2}} \frac{\d^D k_2}{i \pi^{D/2}}
    \frac{1}{(k_1^2 - m_1^2)(k_2^2 - m_2^2)\big((p - k_1 - k_2)^2 - 1\big)}\,,
\end{align}
the closed-form expression in terms of Lauricella $F_C^{(3)}$ reads
\begin{align}\label{eqn:sunsetFC}
    I^{\rm sun}_{1,1,1}
    &=\; -\frac{\Gamma(1-\epsilon)^2 \,\Gamma(\epsilon)\,\Gamma(2\epsilon -1)}{\Gamma(2-\epsilon)}
    \, F_C^{(3)}\!\left(\epsilon,\,2\epsilon -1;\,\epsilon,\epsilon,2-\epsilon;\,z_1,z_2,z_3\right) \nonumber\\
    &\quad
    -\, z_1^{1-\epsilon}\,\frac{\Gamma(1-\epsilon)\,\Gamma(\epsilon -1)\,\Gamma(\epsilon)}{\Gamma(2-\epsilon)}
    \, F_C^{(3)}\!\left(1,\,\epsilon;\,2-\epsilon,\epsilon,2-\epsilon;\,z_1,z_2,z_3\right) \nonumber\\
    &\quad
    -\, z_2^{1-\epsilon}\,\frac{\Gamma(1-\epsilon)\,\Gamma(\epsilon -1)\,\Gamma(\epsilon)}{\Gamma(2-\epsilon)}
    \, F_C^{(3)}\!\left(1,\,\epsilon;\,\epsilon,2-\epsilon,2-\epsilon;\,z_1,z_2,z_3\right) \nonumber\\
    &\quad
    -\, z_1^{1-\epsilon}\, z_2^{1-\epsilon}\,\Gamma(\epsilon -1)^2
    \, F_C^{(3)}\!\left(1,\,2-\epsilon;\,2-\epsilon,2-\epsilon,2-\epsilon;\,z_1,z_2,z_3\right)\,.
\end{align}

We perform a numerical comparison analogous to the previous example. The sunset integral is evaluated numerically using \texttt{AMFlow}, while the Lauricella function representations are computed via \texttt{HyperPrecision} with the prescription \texttt{IDelta} $\to +I$, with a target precision of $30$ correct digits. Sampling 50 points randomly in the parameter space $(m_1^2, m_2^2, p^2)$, we find complete agreement across all sampled points. A representative selection of six points is presented in Table~\ref{tab:sunsetvsFC}.




\setlength{\tabcolsep}{4pt}
\renewcommand{\arraystretch}{1.25}

\begin{table}[ht!]
\centering
\tiny
\caption{%
    Numerical coefficients $c_n$ of the $\epsilon$-expansion $\left( \sum_n c_n\,\epsilon^n \right)$ of Eqs. \eqref{eqn:sunset} and \eqref{eqn:sunsetFC}. Digits shown in red in the \texttt{HyperPrecision} column indicate the trailing positions at which the value differs from the corresponding \texttt{AMFlow} value.
}
\label{tab:sunsetvsFC}
\begin{tabularx}{\textwidth}{c c Y Y}
\toprule
\small{$(m_1^2,m_2^2,p^2)$} & \small{$n$} & \small{\texttt{AMFlow} value of \eeqref{eqn:sunset}} & \small{Numerical Value of the $F_C^{(3)}$ representation of the sunset integral \eeqref{eqn:sunsetFC}} \\
\midrule

\multirow{6}{*}{$\bigl(-\tfrac{10}{9},\;\tfrac{6}{5},\;-\tfrac{7}{8}\bigr)$}
 & $-2$           & $\phantom{-}0.54444444444444444444$
                  & $\phantom{-}0.5444444444444444444$ \\
 & $-1$           & $\phantom{-}1.1218409807965034461 - 3.4906585039886591538\,\mathrm{i}$
                  & $\phantom{-}1.1218409807965034\textcolor{red}{96} - 3.4906585039886591\textcolor{red}{35}\,\mathrm{i}$ \\
 & $\phantom{-}0$ & $\phantom{-}13.1184426812004871162 + 2.1094634019614069985\,\mathrm{i}$
                  & $\phantom{-}13.118442681200487\textcolor{red}{059} + 2.10946340196140\textcolor{red}{7024}\,\mathrm{i}$ \\
 & $\phantom{-}1$ & $\phantom{-}0.1655195941529379681 - 2.4436010853670164358\,\mathrm{i}$
                  & $\phantom{-}0.16551959415293\textcolor{red}{802} - 2.4436010853670164\textcolor{red}{7}\,\mathrm{i}$ \\
 & $\phantom{-}2$ & $\phantom{-}50.558847901283218738 + 2.835376192316706743\,\mathrm{i}$
                  & $\phantom{-}50.558847901283218\textcolor{red}{68} + 2.835376192316706\textcolor{red}{80}\,\mathrm{i}$ \\
 & $\phantom{-}3$ & $\phantom{-}0.2253947434429051901 - 2.6741537958195781239\,\mathrm{i}$
                  & $\phantom{-}0.225394743442905\textcolor{red}{25} - 2.674153795819578\textcolor{red}{20}\,\mathrm{i}$ \\
\midrule

\multirow{6}{*}{$\bigl(\tfrac{10}{3},\;\tfrac{1}{10},\;9\bigr)$}
 & $-2$           & $\phantom{-}2.2166666666666666667$
                  & $\phantom{-}2.2166666666666666667$ \\
 & $-1$           & $-1.9419736195171777557$
                  & $-1.9419736195171777\textcolor{red}{19}$ \\
 & $\phantom{-}0$ & $\phantom{-}17.161567684009542893$
                  & $\phantom{-}17.1615676840095428\textcolor{red}{5}$ \\
 & $\phantom{-}1$ & $-16.331758583797828564$
                  & $-16.331758583797828\textcolor{red}{41}$ \\
 & $\phantom{-}2$ & $\phantom{-}87.868884440849077921$
                  & $\phantom{-}87.868884440849077\textcolor{red}{8}$ \\
 & $\phantom{-}3$ & $-86.756275436396292469$
                  & $-86.756275436396292\textcolor{red}{3}$ \\
\midrule

\multirow{6}{*}{$\bigl(-\tfrac{5}{7},\;2,\;2\bigr)$}
 & $-2$           & $\phantom{-}1.1428571428571428571$
                  & $\phantom{-}1.1428571428571428571$ \\
 & $-1$           & $-0.0174103356242606792 - 2.2439947525641380275\,\mathrm{i}$
                  & $-0.0174103356242606\textcolor{red}{24} - 2.2439947525641380\textcolor{red}{04}\,\mathrm{i}$ \\
 & $\phantom{-}0$ & $\phantom{-}13.5659708596928793150 + 0.1334747813953234749\,\mathrm{i}$
                  & $\phantom{-}13.565970859692879\textcolor{red}{247} + 0.133474781395323\textcolor{red}{511}\,\mathrm{i}$ \\
 & $\phantom{-}1$ & $-1.3511168991270950185 - 1.1266448063309768532\,\mathrm{i}$
                  & $-1.35111689912709\textcolor{red}{497} - 1.126644806330976\textcolor{red}{90}\,\mathrm{i}$ \\
 & $\phantom{-}2$ & $\phantom{-}55.428006034676609753 + 1.147285417134478953\,\mathrm{i}$
                  & $\phantom{-}55.428006034676609\textcolor{red}{68} + 1.14728541713447\textcolor{red}{901}\,\mathrm{i}$ \\
 & $\phantom{-}3$ & $-10.9541149332393771620 - 0.8950510474265994233\,\mathrm{i}$
                  & $-10.9541149332393771 - 0.895051047426599\textcolor{red}{5}\,\mathrm{i}$ \\
\midrule

\multirow{6}{*}{$\bigl(\tfrac{5}{7},\;-2,\;-\tfrac{5}{6}\bigr)$}
 & $-2$           & $-0.14285714285714285714$
                  & $-0.1428571428571428571$ \\
 & $-1$           & $\phantom{-}1.5713123391545282913 - 6.2831853071795864769\,\mathrm{i}$
                  & $\phantom{-}1.571312339154528\textcolor{red}{344} - 6.2831853071795864\textcolor{red}{534}\,\mathrm{i}$ \\
 & $\phantom{-}0$ & $\phantom{-}17.536112153585207860 + 6.922175893860260096\,\mathrm{i}$
                  & $\phantom{-}17.536112153585207\textcolor{red}{784} + 6.922175893860260\textcolor{red}{117}\,\mathrm{i}$ \\
 & $\phantom{-}1$ & $-11.1148504140529442421 - 7.1940334113655109557\,\mathrm{i}$
                  & $-11.114850414052944\textcolor{red}{16} - 7.19403341136551\textcolor{red}{101}\,\mathrm{i}$ \\
 & $\phantom{-}2$ & $\phantom{-}73.573037848917680744 + 7.998834776263532863\,\mathrm{i}$
                  & $\phantom{-}73.573037848917680\textcolor{red}{65} + 7.998834776263532\textcolor{red}{95}\,\mathrm{i}$ \\
 & $\phantom{-}3$ & $-45.873518835293436354 - 8.113559944399110729\,\mathrm{i}$
                  & $-45.873518835293436\textcolor{red}{26} - 8.113559944399110\textcolor{red}{84}\,\mathrm{i}$ \\
\midrule

\multirow{6}{*}{$\bigl(3,\;\tfrac{5}{2},\;\tfrac{9}{5}\bigr)$}
 & $-2$           & $\phantom{-}3.2500000000000000000$
                  & $\phantom{-}3.2500000000000000000$ \\
 & $-1$           & $-0.038465517549680331087$
                  & $-0.0384655175496803\textcolor{red}{28}$ \\
 & $\phantom{-}0$ & $\phantom{-}10.057767779662582565$
                  & $\phantom{-}10.05776777966258256\textcolor{red}{7}$ \\
 & $\phantom{-}1$ & $\phantom{-}10.321812360782795919$
                  & $\phantom{-}10.3218123607827959\textcolor{red}{2}$ \\
 & $\phantom{-}2$ & $\phantom{-}20.742817310512795175$
                  & $\phantom{-}20.74281731051279517$ \\
 & $\phantom{-}3$ & $\phantom{-}71.874303510604452855$
                  & $\phantom{-}71.8743035106044528\textcolor{red}{6}$ \\
\midrule

\multirow{6}{*}{$\bigl(\tfrac{4}{5},\;-\tfrac{7}{4},\;-\tfrac{1}{3}\bigr)$}
 & $-2$           & $\phantom{-}0.025000000000000000000$
                  & $\phantom{-}0.0250000000000000000$ \\
 & $-1$           & $\phantom{-}1.2873150200266141959 - 5.4977871437821381673\,\mathrm{i}$
                  & $\phantom{-}1.287315020026614\textcolor{red}{249} - 5.4977871437821381\textcolor{red}{448}\,\mathrm{i}$ \\
 & $\phantom{-}0$ & $\phantom{-}16.409491626775659156 + 5.056268635257858461\,\mathrm{i}$
                  & $\phantom{-} 16.409491626775659\textcolor{red}{084} + 5.0562686352578584\textcolor{red}{87}\,\mathrm{i}$ \\
 & $\phantom{-}1$ & $-7.4683338855178453061 - 5.1743100394513195164\,\mathrm{i}$
                  & $-7.468333885517845\textcolor{red}{23} - 5.1743100394513195\textcolor{red}{65}\,\mathrm{i}$ \\
 & $\phantom{-}2$ & $\phantom{-}66.150898600686312860 + 5.883255016829411987\,\mathrm{i}$
                  & $\phantom{-}66.150898600686312\textcolor{red}{78} + 5.88325501682941\textcolor{red}{206}\,\mathrm{i}$ \\
 & $\phantom{-}3$ & $-31.161075901005279937 - 5.837019184011361952\,\mathrm{i}$
                  & $-31.161075901005279\textcolor{red}{86} - 5.83701918401136\textcolor{red}{206}\,\mathrm{i}$ \\
\bottomrule
\end{tabularx}
\end{table}

\subsection{Evaluation at a Singular Point}
\label{sec:onsingularity}

Our package can also evaluate MHFs at a singular point by performing an asymptotic
expansion, though the numerical results may be unstable or divergent\footnote{When the target point lies on the singular locus, the package issues a warning to the user.} in some cases, depending on the specific values of the Pochhammer parameters. In the following,
we illustrate cases in which our package yields the correct numerical value at a singularity.

\paragraph{Appell \texorpdfstring{$F_1$}{F1}:}
The following reduction formula of $F_1$ is well known,
\begin{align}\label{f1_red}
    F_1(1; 1, 1; 2; x, x) = \frac{1}{1-x},
\end{align}
which is valid on the singular locus $x = y$ of $F_1$. We checked that our
package yields the correct numerical result for the above case. For example, at $x = 10$, with a requested precision of $15$ digits, we obtain
\begin{align}
    F_1(1; 1, 1; 2; 10, 10) = -0.1111111111111207\ldots,
\end{align}
which agrees with $-1/9$ of Eq.~\eqref{f1_red} to the requested precision.

\paragraph{Appell \texorpdfstring{$F_4$}{F4}:}
The following expressions, in which $F_4$ is evaluated on the boundary
$\sqrt{x} + \sqrt{y} = 1$ of its convergence region, are well known:
\begin{align}
    \frac{4}{\pi} \int_0^{\infty} I_0(t)^2\, K_0(2t)\, dt
    &= F_4\!\left(\tfrac{1}{2}, \tfrac{1}{2};\; 1, 1;\; \tfrac{1}{4}, \tfrac{1}{4}\right) \\
    &= {}_3F_2\!\left(\tfrac{1}{2}, \tfrac{1}{2}, \tfrac{1}{2};\; 1, 1;\; 1\right) \\
    &= {}_2F_1\!\left(\tfrac{1}{4}, \tfrac{1}{4};\; 1;\; 1\right)^{\!2} \\
    &= \frac{\pi}{\Gamma\!\left(\tfrac{3}{4}\right)^{\!4}} = 1.39320392968568\ldots
\end{align}
Note that the point $(x, y) = (1/4, 1/4)$ lies on the principal singular locus
$\sqrt{x} + \sqrt{y} = 1$ of $F_4$, while $z = 1$ is a singularity of both
${}_3F_2$ and ${}_2F_1$. We have verified that with \texttt{HyperPrecision} all expressions above
yield the numerical value $1.39320392968568\ldots$ to the requested precision.

\paragraph{Lauricella \texorpdfstring{$F_C^{(3)}$}{FC(3)}:}
It is well known that integrals of products of Bessel functions yield
Lauricella $F_C^{(N)}$~\cite{Srivastava:1985}. For instance,
\begin{align}
    \int_0^{\infty} e^{-3t}\, I_0(t)^3\, dt
    &= \tfrac{1}{3}\, F_C^{(3)}\!\left[\tfrac{1}{2},\, 1;\; 1, 1, 1;\; \tfrac{1}{9}, \tfrac{1}{9}, \tfrac{1}{9}\right] \\
    &= \bigl(18 + 12\sqrt{2} - 10\sqrt{3} - 7\sqrt{6}\bigr)
       \left\{\frac{2}{\pi}\, K\!\bigl((2-\sqrt{3})(\sqrt{3}-\sqrt{2})\bigr)\right\}^{\!2} \\
    &= 0.50546201971732600605\ldots,
\end{align}
where $K$ denotes the complete elliptic integral of the first kind in the
$K(m)$ convention (i.e.\ $m$ is the parameter, not the modulus). Also, another integral of products of Bessel functions:
\begin{align}
    \frac{6}{\pi} \int_0^{\infty} I_0(t)^3\, K_0(3t)\, dt
    &= F_C^{(3)}\!\left[\tfrac{1}{2}, \tfrac{1}{2};\; 1, 1, 1;\; \tfrac{1}{9}, \tfrac{1}{9}, \tfrac{1}{9}\right] \\
    &= 1.15671541\ldots
\end{align}
Both evaluation points lie on the singular locus of $F_C^{(3)}$ and we have
verified that \texttt{HyperPrecision} can reproduce the numerical result.

\subsection{Evaluation Time}
\label{sec:evatime}

The evaluation time of \texttt{HyperPrecision} is primarily determined by the holonomic rank of the input MHF, the requested order in the $\epsilon$-expansion, and the target numerical precision. The holonomic rank depends on both the number of variables and the structure of the Pochhammer parameters, which together determine the size and analytic complexity of the associated Pfaffian system.

\begin{figure}[!ht]
    \centering
    \includegraphics[width=1\linewidth]{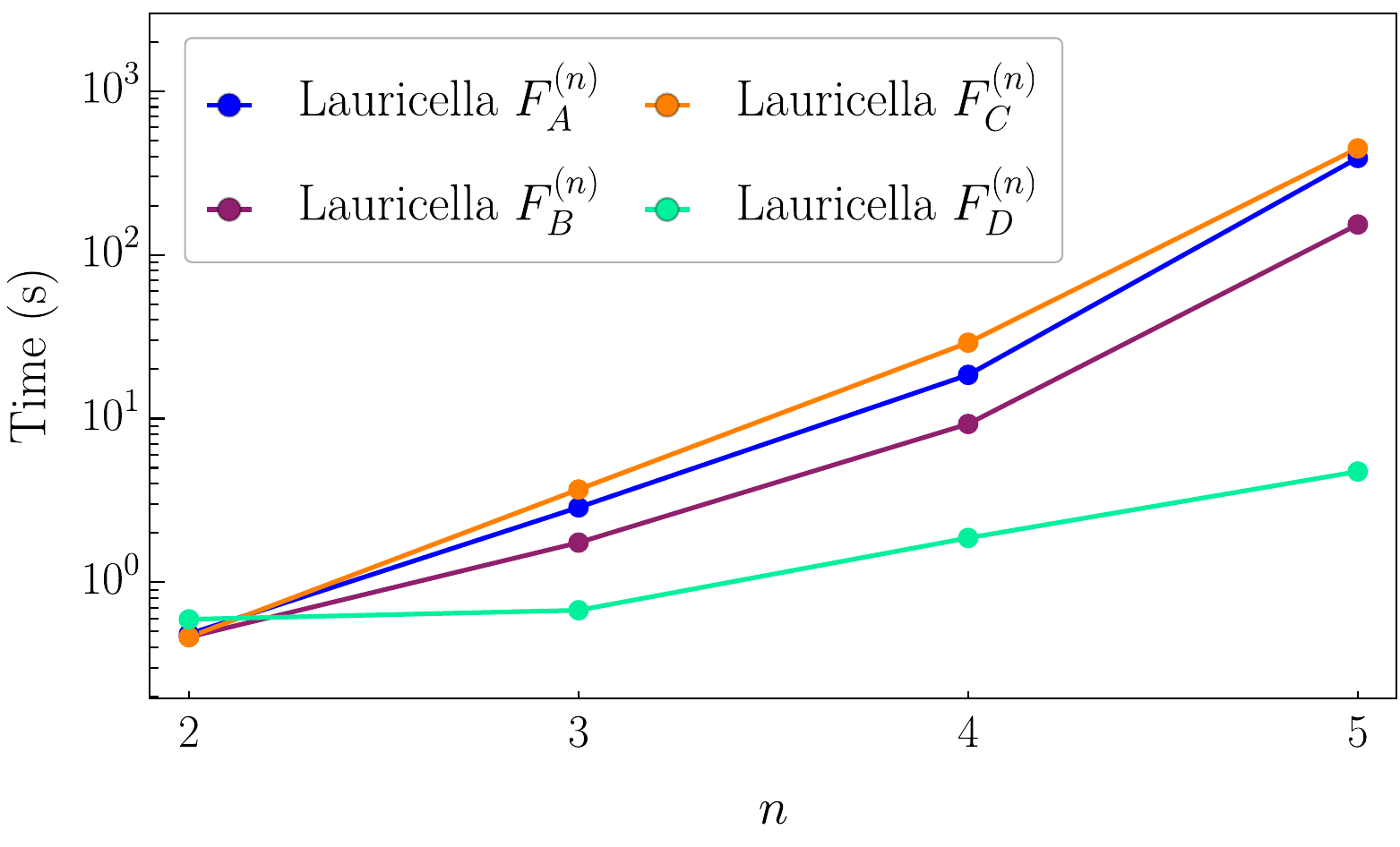}
    \caption{Evaluation time of \texttt{HyperPrecision} for the Lauricella functions $F_A^{(n)}$, $F_B^{(n)}$, $F_C^{(n)}$, and $F_D^{(n)}$ as a function of the number of variables $n$.}
    \label{fig:LCTime}
\end{figure}

\hspace{1cm} To illustrate the typical performance of the package, Figure~\ref{fig:LCTime} shows the evaluation time for the Lauricella functions $F_A^{(n)}$, $F_B^{(n)}$, $F_C^{(n)}$, and $F_D^{(n)}$ with $n=2,3,4,5$, evaluated at randomly chosen target points outside their regions of convergence. In each case, the Laurent expansion was computed from $\mathcal{O}(\epsilon^{-1})$ to $\mathcal{O}(\epsilon^{1})$ with $10$-digit numerical precision. The timings were obtained on an Intel Core i9-14900K processor (24 cores, 32 threads), with the \cmd{ParallelRun} option enabled.

\hspace{1cm} The observed time scaling in Figure~\ref{fig:LCTime} closely follows the growth of the corresponding holonomic ranks. The Lauricella functions $F_A^{(n)}$, $F_B^{(n)}$ and $F_C^{(n)}$ have rank $2^n$, whereas $F_D^{(n)}$ has rank $n+1$, and since the evaluation time grows approximately exponentially with the rank, the timings for $F_A^{(n)}$, $F_B^{(n)}$ and $F_C^{(n)}$ increase much more rapidly with $n$ than those for $F_D^{(n)}$. The remaining time difference among $F_A^{(n)}$, $F_B^{(n)}$ and $F_C^{(n)}$, despite their identical rank, is due to differences in the analytic structure of their Pfaffian systems, in particular to the size of the connection matrices.

\section{Applications}\label{sec:Applications}

Since \texttt{HyperPrecision} can numerically evaluate general Horn-type MHFs, it is applicable to a broad class of problems in theoretical and mathematical physics. In this section, we illustrate this through three examples taken from different research areas: angular integrals in perturbative quantum field theory, cosmological correlators in the cosmological bootstrap, and holographic correlators in non-conformal D$p$-brane holography.

\subsection{Angular Integrals}\label{sec:phasespace}

Phase-space integrals arising in perturbative quantum field theory can be decomposed, after a suitable change of variables, into radial and angular parts~\cite{vanNeerven:1985xr,Beenakker:1988bq,Somogyi:2011ir,Ahmed:2024pxr,Smirnov:2024pbj}. In $D = 4 - 2\epsilon$ space-time dimensions, the angular part with $n$ denominators, in the notation of~\cite{Somogyi:2011ir}, is given by
\begin{equation}\label{angular_n}
\Omega_{j_1, \ldots, j_n} \equiv \int \mathrm{d}\Omega_{D-1}(q) \, \frac{1}{(p_1 \cdot q)^{j_1} \cdots (p_n \cdot q)^{j_n}} \, ,
\end{equation}
where $\{p_i^\mu\}$ are fixed vectors, $\mathrm{d}\Omega_{D-1}(q)$ denotes the rotationally invariant angular measure for the massless vector $q^\mu$, and the powers $\{j_i\}$ depend on the scattering process and the perturbative order under consideration. Below we show how such integrals can be evaluated with \texttt{HyperPrecision}.

\subsubsection{Double-Massive Two-Denominator Integral}

We first consider the double-massive two-denominator angular integral. As shown in~\cite{Lyubovitskij:2021ges}, this integral can be expressed in terms of the Lauricella function $F_B^{(3)}$ as
\begin{align}
    \label{eqn:angularFB}
    \Omega_{j,\ell}(v_1,v_2)
    \;=\;
    \frac{2\pi\, v_{12}^{1-j-\ell-\varepsilon}}{1-2\varepsilon}\,
    F_B^{(3)}\!\left(
        \tfrac{j}{2},\tfrac{\ell}{2},\tfrac{3-j-\ell}{2}-\varepsilon;\,
        \tfrac{j+1}{2},\tfrac{\ell+1}{2},\tfrac{1-j-\ell}{2}-\varepsilon;\,
        \tfrac{3}{2}-\varepsilon;\,
        x_1,x_2,x_3
    \right),
\end{align}
where
\begin{align}
    \label{eqn:angularxsub}
    x_1 = 1-\frac{v_{11}}{v_{12}},\qquad
    x_2 = 1-\frac{v_{22}}{v_{12}},\qquad
    x_3 = 1-v_{12},
\end{align}
and $v_{ij}=v_i\cdot v_j$, where $v_i$'s are fixed vectors in $d = 4 - 2\varepsilon$ dimensional Minkowski space.

\hspace{1cm} For illustration, we take unit propagator powers, $j=\ell=1$, and evaluate the integral at the randomly chosen kinematic point 
\begin{align}
    \label{eqn:angularvsub}
    v_{11} = \tfrac{1}{3},\qquad
    v_{12} = \tfrac{28}{5},\qquad
    v_{22} = \tfrac{21}{11}.
\end{align}
This corresponds to $(x_1,x_2,x_3) = (79/84,\,29/44,\,-23/5)$, which lies outside the convergence region $|x_1|<1$, $|x_2|<1$, $|x_3|<1$ of the defining series of $F_B^{(3)}$. Nevertheless, using \cmd{HypFunctionExpand[]}, we can obtain the $\varepsilon$-expansion easily as follows:
\begin{lstlisting}[language=Mathematica,literate={->}{{$\to$}}1 {\\[Epsilon]}{{$\varepsilon$}}1 {\\[Pi]}{{$\pi$}}1]
In[]:=  HypVal = HypFunctionExpand[LauricellaFB[{1/2, 1/2, 1/2 - \[Epsilon], 1, 1, -\[Epsilon]},
            {3/2 - \[Epsilon]}, {79/84, 29/44, -(23/5)}], {\[Epsilon], 4}, 30];

In[]:=  Series[(2 \[Pi])/(1 - 2 \[Epsilon]) v12^(1 - j - l - \[Epsilon]) * HypVal
            /. {v12 -> 28/5, j -> 1, l -> 1}, {\[Epsilon], 0, 4}] // Normal // N[#, 30] &

Out[]=  2.988957928338904937639407592 + 5.026237915393715670571844297 \[Epsilon]
        + 9.723950427800003063380764239 \[Epsilon]^2
        + 19.32428880508757066003927207 \[Epsilon]^3
        + 38.60064440264985739801732964 \[Epsilon]^4
\end{lstlisting}
We find perfect agreement with the analytic expression from Eq.~(3.194) of~\cite{Lyubovitskij:2021ges} through $\mathcal{O}(\varepsilon)$,
\begin{align}
    \label{eqn:angular11analytic}
    \Omega_{1,1}\big|_{j=\ell=1}
    &=
    \frac{\pi}{\sqrt{X}}\!\left[
        \log\!\left(\frac{v_{12}+\sqrt{X}}{v_{12}-\sqrt{X}}\right)
        - \varepsilon\!\left(
            \tfrac{1}{2}\log^2\!\frac{v_{11}}{v_{13}^{2}}
          - \tfrac{1}{2}\log^2\!\frac{v_{22}}{v_{23}^{2}}
          \right.\right.
    \nonumber\\
    &\hspace{6.5em}
        + 2\,\mathrm{Li}_2\!\left(1-\frac{v_{13}}{1-\sqrt{1-v_{11}}}\right)
        + 2\,\mathrm{Li}_2\!\left(1-\frac{v_{13}}{1+\sqrt{1-v_{11}}}\right)
    \nonumber\\
    &\hspace{6.5em}
        \left.\left.
        - 2\,\mathrm{Li}_2\!\left(1-\frac{v_{23}}{1-\sqrt{1-v_{22}}}\right)
        - 2\,\mathrm{Li}_2\!\left(1-\frac{v_{23}}{1+\sqrt{1-v_{22}}}\right)
        \right)
    \right] + \mathcal{O}(\varepsilon^2)\,,
\end{align}
where $X = v_{12}^2 - v_{11}v_{22}$ and
\begin{align}
    v_{13} = \frac{v_{11}(v_{22}+\sqrt{X}) - v_{12}(v_{12}+\sqrt{X})}{v_{11}+v_{22}-2v_{12}},
    \qquad
    v_{23} = \frac{v_{22}(v_{11}-\sqrt{X}) - v_{12}(v_{12}-\sqrt{X})}{v_{11}+v_{22}-2v_{12}}\,.
\end{align}

\subsubsection{Massless Three-Denominator Integral}\label{subsec:angular3prop}

We next turn to the massless case, $p_i^2=0$, with three denominators and unit propagator powers.
The corresponding angular integral, $\Omega_{1,1,1}(v_{12},v_{13},v_{23})$, was first computed through $\mathcal{O}(\epsilon^0)$ in~\cite{Somogyi:2011ir} and later calculated analytically to $\mathcal{O}(\epsilon^1)$ and $\mathcal{O}(\epsilon^2)$ in~\cite{Haug:2024yfi} and~\cite{Ahmed:2024pxr}, respectively. An explicit analytic result to all orders in $\epsilon$ appeared in~\cite{Haug:2025sre}.
To evaluate it numerically with \texttt{HyperPrecision}, we start from its MB representation~\cite{Somogyi:2011ir},
\begin{align}\label{eqn:omega111MB}
    \Omega_{1,1,1}(v_{12},v_{13},v_{23},\epsilon)
    &=
    \frac{2^{-1-2\epsilon}\,\pi^{1-\epsilon}}{\Gamma(-1-2\epsilon)}
    \int \frac{\d z_{12}\,\d z_{13}\,\d z_{23}}{(2\pi\ic)^{3}}\,
    \Gamma(-z_{12})\,\Gamma(-z_{13})\,\Gamma(-z_{23})\,\Gamma(-2-\epsilon-z)
    \notag\\
    &\quad\times
    \Gamma(1+z_{12}+z_{13})\,\Gamma(1+z_{12}+z_{23})\,\Gamma(1+z_{13}+z_{23})\,
    v_{12}^{\,z_{12}}\,v_{13}^{\,z_{13}}\,v_{23}^{\,z_{23}},
\end{align}
with $z \equiv z_{12}+z_{13}+z_{23}$. Unlike the two-denominator example above, to our knowledge, this integral cannot be written as a single Horn-type MHF. Instead, using \texttt{MBConicHulls}~\cite{Ananthanarayan:2020fhl}, the MB integral can be written as a sum of five MHFs,
\begin{align}\label{eqn:omega111SeriesSum}
    \Omega_{1,1,1}(v_{12},v_{13},v_{23},\epsilon)
    \;=\;
    \mathcal{N}(\epsilon)\,\sum_{k=1}^{5}\,S_{k},
\end{align}
where
\begin{align}
    \mathcal{N}(\epsilon)
    \;\equiv\;
    \frac{2^{-1-2\epsilon}\,\pi^{1-\epsilon}}{\Gamma(-1-2\epsilon)}.
\end{align}
Each $S_k$ has the common form
\begin{align}\label{eqn:Sktemplate}
    S_{k}
    \;=\;
    \sum_{n_{1},n_{2},n_{3}=0}^{\infty}
    \frac{(-1)^{n_{1}+n_{2}+n_{3}}}{n_{1}!\,n_{2}!\,n_{3}!}\,
    \mathcal{T}_{k}(n_{1},n_{2},n_{3};\epsilon),
\end{align}
with the five summands
\begin{align}
    \mathcal{T}_{1}
    &=
    v_{12}^{\,n_{1}}\,v_{13}^{\,n_{2}}\,v_{23}^{\,n_{3}}\,
    \Gamma(1+n_{1}+n_{2})\,\Gamma(1+n_{1}+n_{3})\,\Gamma(1+n_{2}+n_{3})
    \notag\\
    &\quad\times
    \Gamma(-2-\epsilon-n_{1}-n_{2}-n_{3}), \label{eqn:T1}\\[4pt]
    \mathcal{T}_{2}
    &=
    v_{12}^{\,n_{1}}\,v_{13}^{\,n_{2}}\,v_{23}^{\,-2-\epsilon-n_{1}-n_{2}+n_{3}}\,
    \Gamma(1+n_{1}+n_{2})\,\Gamma(2+\epsilon+n_{1}+n_{2}-n_{3})
    \notag\\
    &\quad\times
    \Gamma(-1-\epsilon-n_{1}+n_{3})\,\Gamma(-1-\epsilon-n_{2}+n_{3}), \label{eqn:T2}\\[4pt]
    \mathcal{T}_{3}
    &=
    v_{12}^{\,n_{1}}\,v_{13}^{\,-1-\epsilon+n_{2}+n_{3}}\,v_{23}^{\,-1-n_{1}-n_{2}}\,
    \Gamma(1+n_{1}+n_{2})\,\Gamma(1+\epsilon-n_{2}-n_{3})
    \notag\\
    &\quad\times
    \Gamma(-1-\epsilon-n_{1}+n_{3})\,\Gamma(-\epsilon+n_{1}+n_{2}+n_{3}), \label{eqn:T3}\\[4pt]
    \mathcal{T}_{4}
    &=
    v_{12}^{\,-1-\epsilon+n_{2}+n_{3}}\,v_{13}^{\,n_{1}}\,v_{23}^{\,-1-n_{1}-n_{2}}\,
    \Gamma(1+n_{1}+n_{2})\,\Gamma(1+\epsilon-n_{2}-n_{3})
    \notag\\
    &\quad\times
    \Gamma(-1-\epsilon-n_{1}+n_{3})\,\Gamma(-\epsilon+n_{1}+n_{2}+n_{3}), \label{eqn:T4}\\[4pt]
    \mathcal{T}_{5}
    &=
    v_{12}^{\,-1-\epsilon+n_{2}+n_{3}}\,v_{13}^{\,-1-\epsilon+n_{1}+n_{3}}\,v_{23}^{\,\epsilon-n_{1}-n_{2}-n_{3}}\,
    \Gamma(1+\epsilon-n_{1}-n_{3})\,\Gamma(1+\epsilon-n_{2}-n_{3})
    \notag\\
    &\quad\times
    \Gamma(-\epsilon+n_{1}+n_{2}+n_{3})\,\Gamma(-1-2\epsilon+n_{1}+n_{2}+2n_{3}). \label{eqn:T5}
\end{align}

\hspace{1cm} Since all five series $S_{k}$ are Horn-type triple hypergeometric series of finite holonomic rank, their Laurent expansion can be evaluated one by one with the command \cmd{HypExpand[]}, and then the expansion of $\Omega_{1,1,1}$ can be obtained using Eq.~\eqref{eqn:omega111SeriesSum}. We have verified this for several randomly chosen kinematic points $(v_{12},v_{13},v_{23})$, finding complete numerical agreement with the analytic expression provided in the ancillary file \texttt{3prop0mass.m} of~\cite{Ahmed:2024pxr} up to $\mathcal{O}(\epsilon^{2})$, the highest order provided in that reference.

\hspace{1cm}
We include this example in the ancillary notebook \texttt{Examples.nb}, where we demonstrate how to numerically compute $\Omega_{1,1,1}$ up to $\mathcal{O}(\epsilon^{10})$ with 50-digit precision. We have verified that our result agrees numerically up to $\mathcal{O}(\epsilon^{2})$ with~\cite{Ahmed:2024pxr} and up to $\mathcal{O}(\epsilon^{10})$ with~\cite{Haug:2025sre}.

\subsection{Cosmological Correlators}\label{sec:correlators}

Cosmological correlators encode the statistics of primordial fluctuations generated during inflation and play a central role in the cosmological bootstrap programme~\cite{Arkani-Hamed:2018kmz,Baumann:2022jpr}. Recent progress has led to closed-form expressions for correlators involving massive-field exchange. In particular, the results of~\cite{Aoki:2024uyi} express three-point functions with double-massive exchanges in terms of MHFs evaluated outside their region of convergence, making them a natural testing ground for \texttt{HyperPrecision}.

\hspace{1cm} In the following, we consider the correlators studied in~\cite{Aoki:2024uyi}, where the seed integrals $I^{p_1p_2p_3}_{abc,\alpha\beta}$ were obtained in closed form as the sum of an Appell $F_4$ function and an infinite series of one-variable hypergeometric functions, given in Eqs.~(3.42)--(3.44) of ~\cite{Aoki:2024uyi}. The explicit expressions are presented in Appendix \ref{app:seed_ints} for completeness. Here, $a,b,c=\pm$ denote Schwinger--Keldysh indices, while $\alpha,\beta$ label the exchanged massive fields.
The connection with cosmological observables is made through the combination
\begin{align}\label{eqn:Sigma}
    \Sigma(u,v) = 2\,\mathrm{Re}\!\left[
        I^{+-+}(u,v) + I^{++-}(u,v) + I^{-++}(u,v) + I^{+++}(u,v)
    \right],
\end{align}
from which the normalised bispectrum shape function follows~\cite{Aoki:2024uyi},
\begin{align}\label{eqn:shape}
    \frac{S(k,x_2 k,x_3 k)}{S(k,k,k)}
    =
    \frac{1}{\Sigma(1,1)}\!\left[
        \frac{x_3}{x_2^2}\,\Sigma\!\left(\tfrac{1}{x_2},\tfrac{x_3}{x_2}\right)
        + x_2 x_3\,\Sigma(x_2,x_3)
        + \frac{x_2}{x_3^2}\,\Sigma\!\left(\tfrac{1}{x_3},\tfrac{x_2}{x_3}\right)
    \right],
\end{align}
where $x_2$ and $x_3$ are the dimensionless momentum ratios. We focus on the case $p_1=p_2=p_3=-2$ with mass parameter $\mu=2$, corresponding to $m=5H/2$, for which a comparison with independent methods is available in~\cite{Aoki:2024uyi}.

\begin{figure}[!ht]
    \centering
    \includegraphics[width=1\linewidth]{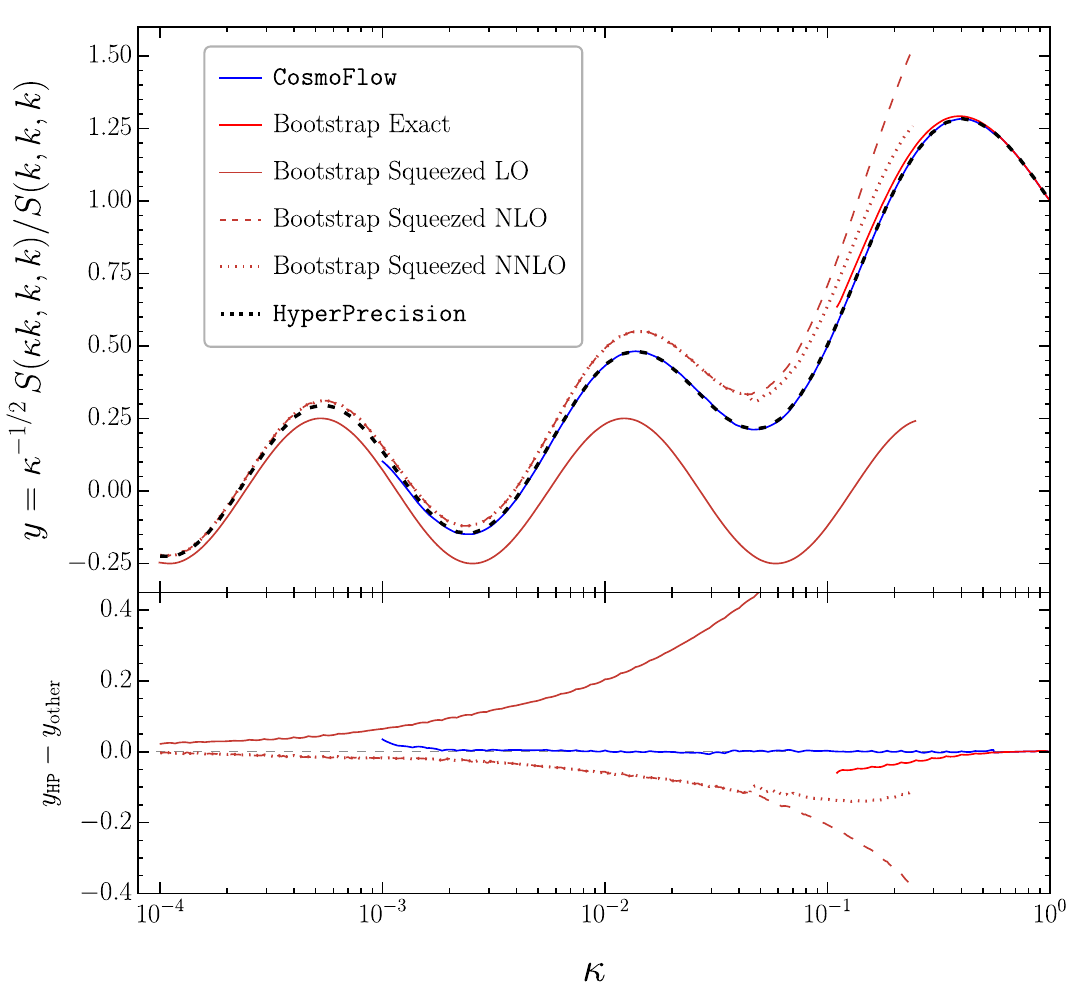}
    \caption{The normalised bispectrum shape function $\kappa^{-1/2}\, S(\kappa k,k,k)/S(k,k,k)$ at $\mu=2$ ($m=5H/2$), evaluated with \texttt{HyperPrecision} (black dashed) and compared with \texttt{CosmoFlow}~\cite{Werth:2024aui} (blue), the bootstrap-exact result of~\cite{Aoki:2024uyi} (orange), and its squeezed-limit LO, NLO, and NNLO expansions (red). The lower panel shows the pointwise difference between \texttt{HyperPrecision} and each of the comparison curves. The plot reproduces the third panel of Figure~7 of~\cite{Aoki:2024uyi}.}
    \label{fig:shape2D}
\end{figure}
\hspace{1cm} The $F_4(\dots, u^2, v^2)$ contribution, in the seed integrals,  converges only for $u+v<1$ in the kinematic variables $u=k_1/k_{2}$ and $v=k_3/k_{2}$, whereas the physical region of the three-point function lies outside this domain. Reaching this region in~\cite{Aoki:2024uyi} required a delicate analytic continuation in which apparent divergences of Eqs.~(3.42)-(3.44) of ~\cite{Aoki:2024uyi} cancel. An alternative route is to rewrite the one-variable hypergeometric series in Eqs.~(3.42)-(3.44) as a genuine two-variable hypergeometric series. Using the \texttt{Olsson} package~\cite{Ananthanarayan:2021yar} to manipulate the bivariate series, the normalised bispectrum shape function can be expressed as a sum of $27$ bivariate hypergeometric series of Appell $F_4$ and Kamp\'e de F\'eriet type. The analytic continuation of each of these series outside its convergence region is then handled automatically by \texttt{HyperPrecision}. In this way, the normalised bispectrum shape function can be evaluated directly in the physical region without any further special function manipulation.

\hspace{1cm} In Figure~\ref{fig:shape2D}, we plot the normalised shape function $\kappa^{-1/2}\,S(\kappa k,k,k)/S(k,k,k)$ as a function of $\kappa$, comparing the \texttt{HyperPrecision} evaluation of Eq.~\eqref{eqn:shape} with independent results and reproducing the third panel of Figure~7 of~\cite{Aoki:2024uyi}. The squeezed-limit expansions are accurate only for $\kappa \ll 1$.
The bootstrap-exact result is valid throughout the displayed range, but requires the explicit analytic continuation.
In contrast, the \texttt{HyperPrecision} result agrees with the bootstrap-exact curve and with \texttt{CosmoFlow}~\cite{Werth:2024aui} across the full kinematic range, without any additional analytic continuation step.

\begin{figure}[!ht]
    \centering
    \includegraphics[width=1\linewidth]{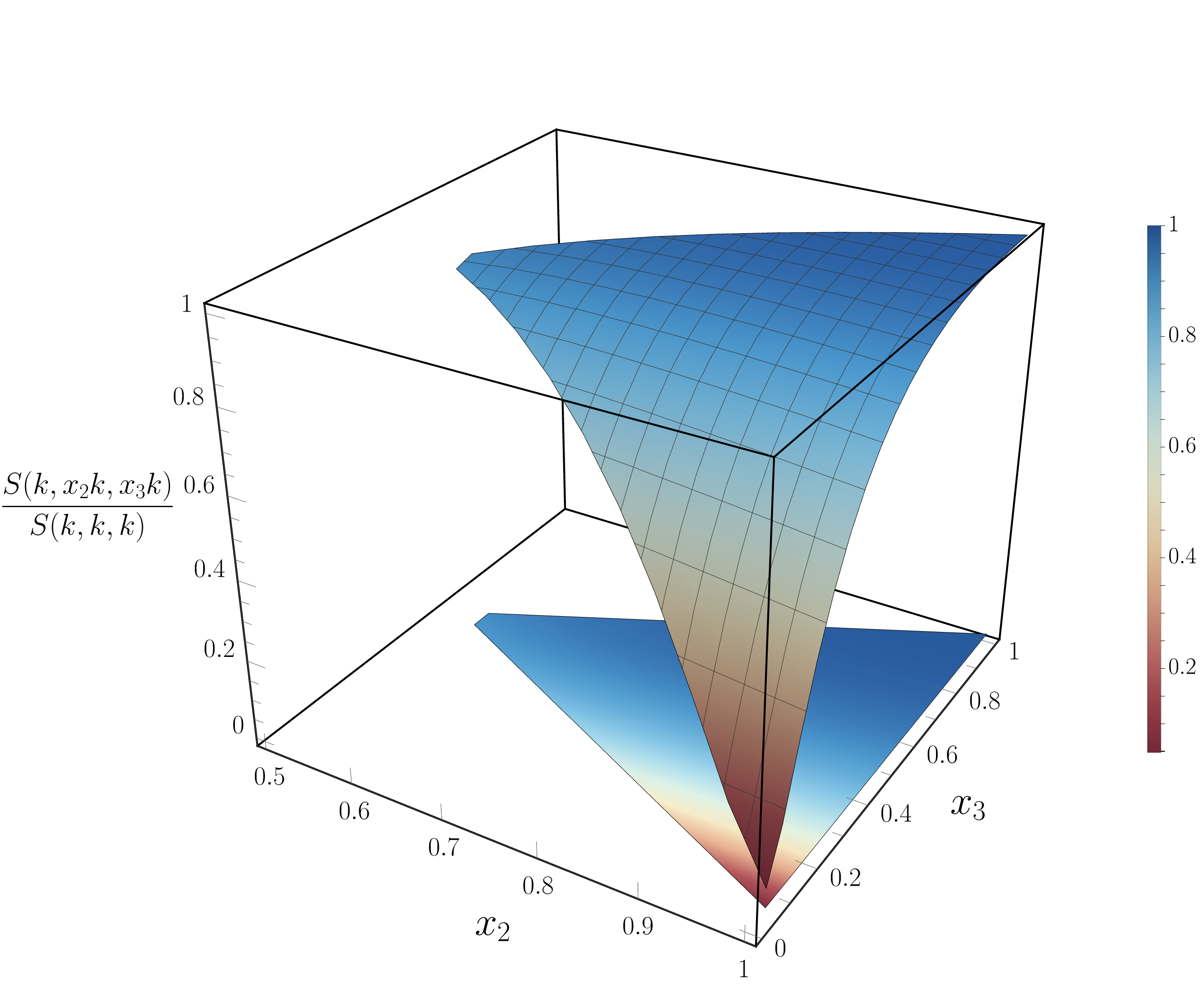}
    \caption{The normalised bispectrum shape function $S(k,x_2 k,x_3 k)/S(k,k,k)$ at $\mu=2$ ($m=5H/2$), shown as a surface over the physical kinematic region in the $(x_2,x_3)$ plane and evaluated with \texttt{HyperPrecision}. This is analogous to Figure~5 of~\cite{Aoki:2024uyi}, which displays the same shape function at $m=2H$.}
    \label{fig:shape3D}
\end{figure}

\hspace{1cm} Finally, the full two-dimensional kinematic dependence of the shape function, evaluated with \texttt{HyperPrecision} over the physical region of the $(x_2,x_3)$ plane, is shown in Figure~\ref{fig:shape3D}. As the MHF representation of the shape function is not valid for $x_2 + x_3 \simeq 1$, we plot the shape function for $x_2 \in [0.55,1]$ and $x_3 \ge 1.05-x_2$. Its qualitative features agree with those of Figure~5 of~\cite{Aoki:2024uyi}, and we checked that it numerically matches with \texttt{CosmoFlow} within $0.03 \%$ accuracy. The \texttt{HyperPrecision} implementation of this example is provided in the ancillary notebook \texttt{Examples.nb}. For further details on the cosmological setup, we refer the reader to~\cite{Aoki:2024uyi}.

\subsection{Holographic Correlators}\label{sec:holographic}

As a third application, we consider three-point correlators of scalar operators in non-conformal D$p$-brane holography, recently studied in~\cite{Bobev:2025idz}. In that work, these correlators were computed in the large-$N$ and strong-coupling limit of the $(p+1)$-dimensional maximally supersymmetric Yang--Mills theory dual to the D$p$-brane background, using an auxiliary AdS structure of fractional dimension. In the notation of~\cite{Bobev:2025idz}, the resulting three-point function of scalar operators with conformal dimensions $\Delta_i$ can be written as
\begin{small}
\begin{align}\label{eqn:holo3pt}
&\langle\mathcal{O}_1(x)\,\mathcal{O}_2(y)\,\mathcal{O}_3(z)\rangle_\eta
    = \mathcal{N}_{\Delta_i}\,(d_{yz}^2)^{\eta-\Theta}\bigg[
       \Gamma(\tfrac{\eta}{2}-\Theta_2)\Gamma(\tfrac{\eta}{2}-\Theta_3)\Gamma(\Theta-\eta)\Gamma(\Theta_2+\Theta_3-\tfrac{\eta}{2}) \notag\\
    &\hspace{2em}\times F_4\!\left(\Theta-\eta,\Theta_2+\Theta_3-\tfrac{\eta}{2};1+\Theta_3-\tfrac{\eta}{2},1+\Theta_2-\tfrac{\eta}{2};A,B\right) \notag\\
    &\quad + B^{\eta/2-\Theta_2}\Gamma(\Theta_2-\tfrac{\eta}{2})\Gamma(\tfrac{\eta}{2}-\Theta_3)\Gamma(\Theta_3)\Gamma(\Theta_1+\Theta_3-\tfrac{\eta}{2})\,F_4\!\left(\Theta_3,\Theta_1+\Theta_3-\tfrac{\eta}{2};1+\Theta_3-\tfrac{\eta}{2},1-\Theta_2+\tfrac{\eta}{2};A,B\right) \notag\\
    &\quad + A^{\eta/2-\Theta_3}\Gamma(\tfrac{\eta}{2}-\Theta_2)\Gamma(\Theta_3-\tfrac{\eta}{2})\Gamma(\Theta_2)\Gamma(\Theta_1+\Theta_2-\tfrac{\eta}{2})\,F_4\!\left(\Theta_2,\Theta_1+\Theta_2-\tfrac{\eta}{2};1-\Theta_3+\tfrac{\eta}{2},1+\Theta_2-\tfrac{\eta}{2};A,B\right) \notag\\
    &\quad + A^{\eta/2-\Theta_3}B^{\eta/2-\Theta_2}\Gamma(\Theta_2-\tfrac{\eta}{2})\Gamma(\Theta_3-\tfrac{\eta}{2})\Gamma(\Theta_1)\Gamma(\tfrac{\eta}{2})\,F_4\!\left(\Theta_1,\tfrac{\eta}{2};1-\Theta_3+\tfrac{\eta}{2},1-\Theta_2+\tfrac{\eta}{2};A,B\right)\bigg],
\end{align}
\end{small}
where $\Theta_i = \tfrac{1}{2}(\Delta_1+\Delta_2+\Delta_3)-\Delta_i$, $\Theta = \Theta_1+\Theta_2+\Theta_3$, and the normalisation factor is $\mathcal{N}_{\Delta_i} = \pi^\eta/(\Gamma(\Theta_1)\Gamma(\Theta_2)\Gamma(\Theta_3)\Gamma(\eta/2))$. The kinematic variables $A = d_{xy}^2/d_{yz}^2$ and $B = d_{xz}^2/d_{yz}^2$ are built from the Euclidean distances $d_{ij} = |x_i - x_j|$, while $\eta = (3-p)^2/(5-p)$ encodes the non-conformality of the D$p$-brane background.

\hspace{1cm} The main difficulty in evaluating Eq.~\eqref{eqn:holo3pt} numerically is that the physical region lies outside the convergence domain of the Appell $F_4$ series. Indeed, as pointed out in~\cite{Bobev:2025idz}, the triangle inequality $d_{xy}+d_{xz}\geq d_{yz}$ implies
$\sqrt{A}+\sqrt{B}\geq 1$, whereas the defining series of the Appell $F_4$ functions converge only for $\sqrt{A}+\sqrt{B}<1$. Therefore, numerical evaluation in the physical region requires a non-trivial analytic continuation of $F_4$, which was derived explicitly in~\cite{Bobev:2025idz}. With \texttt{HyperPrecision}, this continuation is performed automatically through the associated Pfaffian system. As a result, Eq.~\eqref{eqn:holo3pt} can be evaluated directly in the physical region without carrying out the analytic continuation by hand.

\begin{figure}[ht!]
    \centering
    \begin{subfigure}[b]{1\linewidth}
        \centering
        \includegraphics[width=\linewidth]{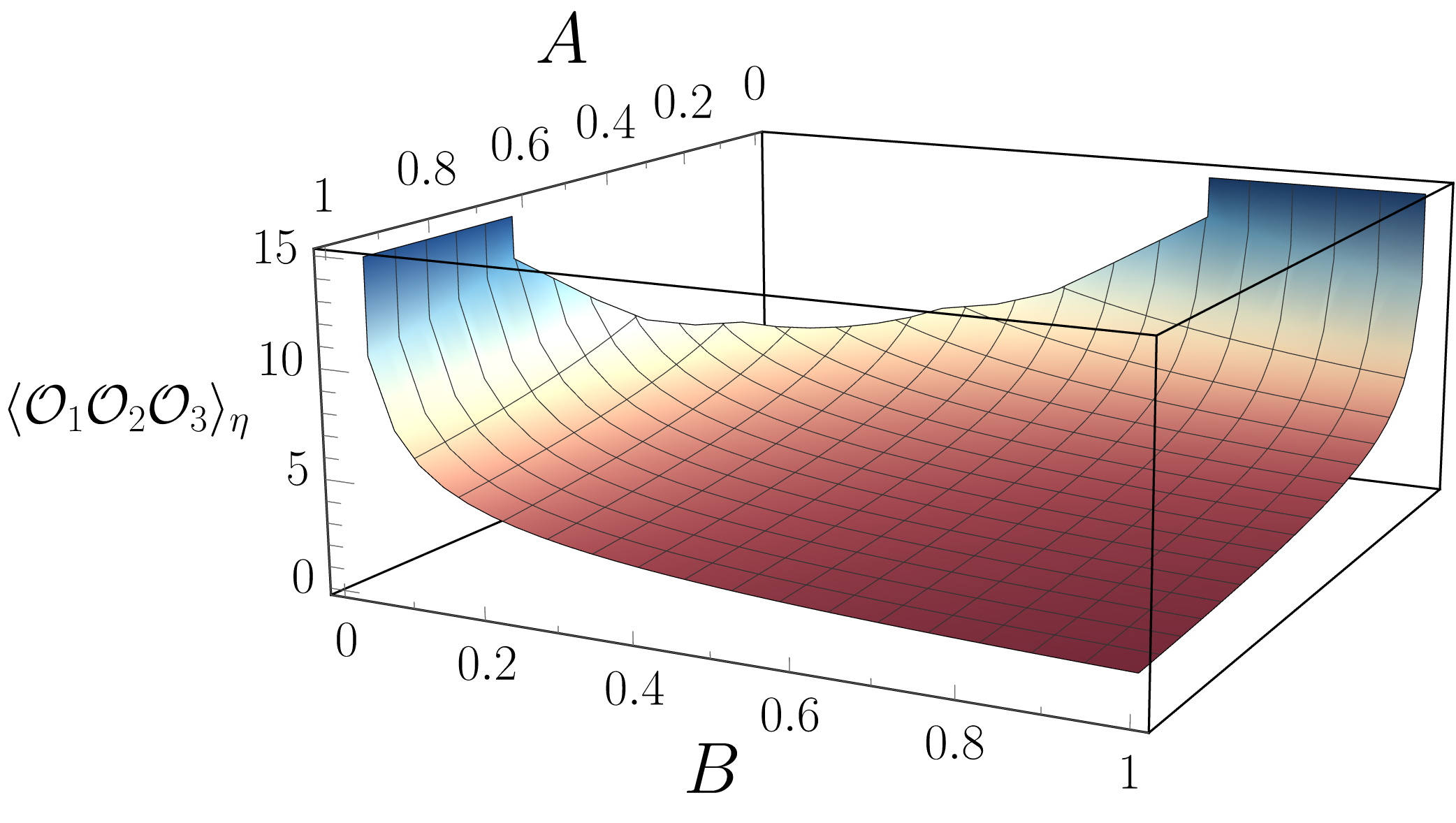}
        \caption{$p=2$, $\ell_i = (2,2,2)$.}
        \label{fig:3pt-222}
    \end{subfigure}

    \par\vspace{1cm}

    \begin{subfigure}[b]{1\linewidth}
        \centering
        \includegraphics[width=\linewidth]{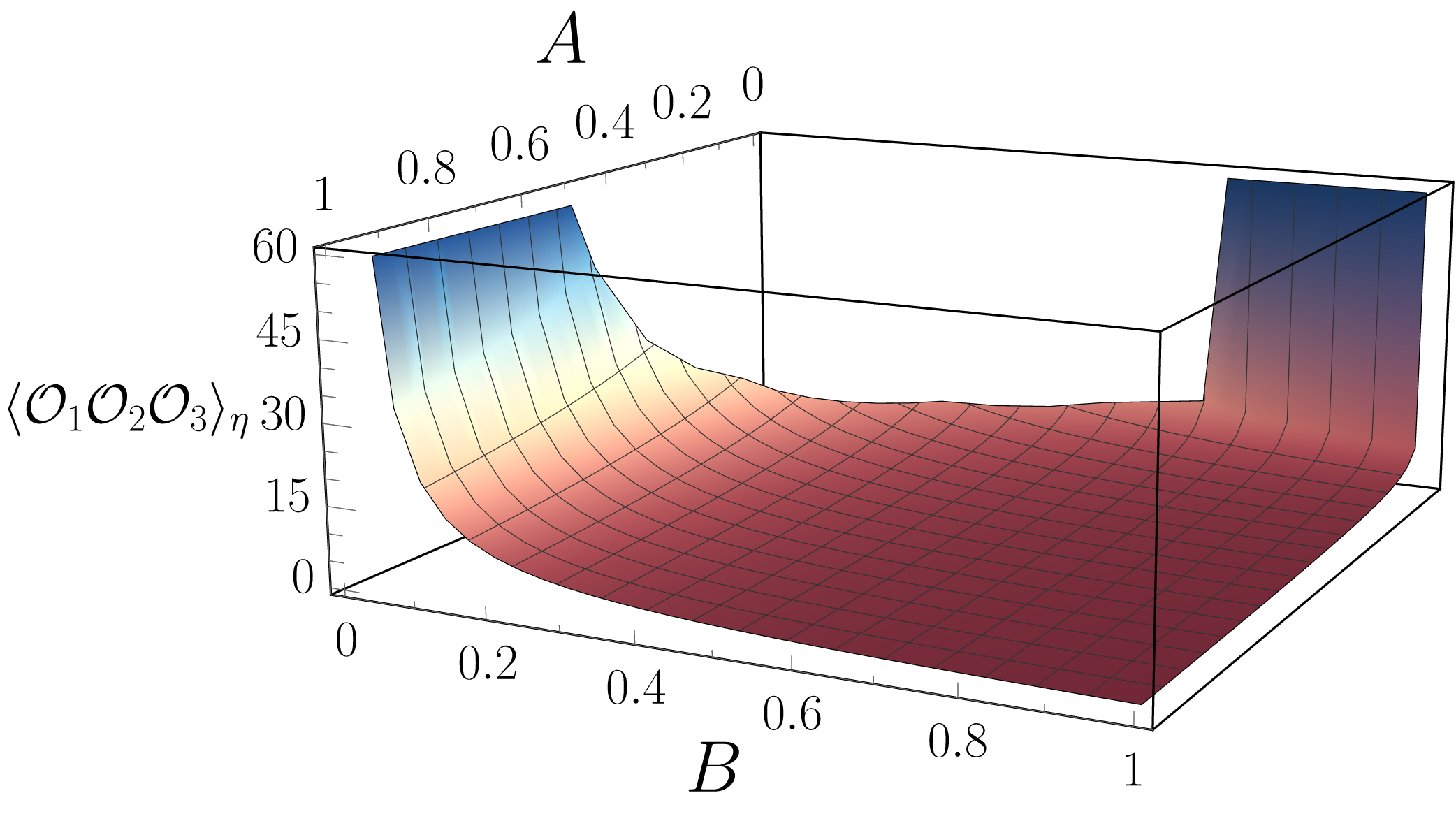}
        \caption{$p=2$, $\ell_i = (3,3,4)$.}
        \label{fig:3pt-334}
    \end{subfigure}
    \caption{The analytically continued three-point function $\langle\mathcal{O}_1\mathcal{O}_2\mathcal{O}_3\rangle_\eta$ as a function of $A$ and $B$ for $p=2$ and two choices of $\ell_i$. The plots are restricted to the physical region $\sqrt{A}+\sqrt{B}\geq 1$ within the unit square, as discussed in~\cite{Bobev:2025idz}.}
    \label{fig:3pt-p2}
\end{figure}
\hspace{1cm} We illustrate this in Figure~\ref{fig:3pt-p2} for $p=2$ and two choices of $\ell_i$, namely $(2,2,2)$ and $(3,3,4)$. The corresponding conformal dimensions are
$\Delta_i = \tfrac{7-p}{5-p}(b+2) + \tfrac{2}{5-p}\ell_i$,
with $b=-2$. We further set $d_{yz}=1$ and use the regularisation $V_\eta=1$ for the infinite volume factor, following~\cite{Bobev:2025idz}. The resulting plots agree with Figure~2 of~\cite{Bobev:2025idz}, including the characteristic divergent behaviour at the corners $(A,B)=(0,1)$ and $(1,0)$, corresponding to the coincident-point limits $x\to y$ and $x\to z$, respectively.

\hspace{1cm} The \texttt{HyperPrecision} implementation of this example is also provided in the ancillary notebook \texttt{Examples.nb}. For further details on the holographic setup, we refer the reader to~\cite{Bobev:2025idz}.

\section{Summary and Discussions}\label{sec:Conclusion}

Multivariate hypergeometric functions arise in many areas of mathematics and physics, from algebraic geometry and number theory to scattering amplitudes, cosmological correlators, and holographic correlation functions. Despite their widespread appearance, their high-precision numerical evaluation remains difficult in practice. The main difficulty is that their defining series converge only in limited regions of argument space, while analytic continuation to arguments of interest is, in general, highly non-trivial.

\hspace{1cm} In this paper, we have presented \texttt{HyperPrecision}, a publicly available \textsc{Mathematica} package developed to numerically compute a broad class of complete Horn-type MHFs with real arguments. Given an input function, the package constructs the associated Pfaffian system dynamically using \texttt{FiniteFlow}. It then restricts this system to a one-dimensional contour and solves the resulting ordinary differential equation with \texttt{DESolver} distributed with the \texttt{AMFlow} package~\cite{Liu:2022chg}. When the function depends on a small parameter $\epsilon$, the Laurent expansion is reconstructed by interpolation from evaluations on a finite grid of $\epsilon$ values. The package is accompanied by an ancillary notebook \texttt{Examples.nb}, which illustrates its usage and provides complete implementations of the examples discussed in this paper, along with several additional examples.

\hspace{1cm} We have tested the package on MHFs of up to six variables, including the Appell functions, Horn $G$- and $H$-series, and the Lauricella $F_A$, $F_B$, $F_C$ and $F_D$ functions, as discussed in Section~\ref{sec:validation}. In all cases considered, we find numerical agreement with existing public codes and known analytic results. We also showed in Section~\ref{sec:onsingularity} that, although the package is primarily designed for regular target points, it can produce correct values at certain singular points through asymptotic expansions, as illustrated for Appell $F_1$, Appell $F_4$ and Lauricella $F_C^{(3)}$.

\hspace{1cm} Beyond these tests, we have also illustrated the broader applicability of \texttt{HyperPrecision} with three applications from high-energy physics, presented in Section~\ref{sec:Applications}. For angular integrals, we evaluated the massless three-denominator integral $\Omega_{1,1,1}$ up to $\mathcal{O}(\epsilon^{10})$ with $50$-digit precision. For double-massive-exchange cosmological correlators, we evaluated the normalised bispectrum shape function directly in the physical region, reproducing both the bootstrap-exact result of~\cite{Aoki:2024uyi} and the \texttt{CosmoFlow}~\cite{Werth:2024aui} evaluation, without using the explicit analytic continuation derived in~\cite{Aoki:2024uyi}. Finally, for three-point holographic correlators in non-conformal D$p$-brane backgrounds, we evaluated the analytically continued correlator of~\cite{Bobev:2025idz} directly in the physical region, bypassing the manual analytic continuation of the Appell $F_4$ functions used in that reference.

\hspace{1cm} While the present implementation already covers a broad class of examples, there are several directions in which \texttt{HyperPrecision} can be improved, both to increase its efficiency and to extend the class of functions and kinematic configurations it can handle. We outline some of these directions below and leave them for future work.

\begin{enumerate}
    \item The holonomic rank of an MHF can drop for special values of its parameters or arguments. Such degenerate cases are not yet treated automatically in the current version of the package, and incorporating them would allow \texttt{HyperPrecision} to handle a broader class of higher-dimensional MHFs such as those appearing in high-energy physics.

    \item The dominant cost of the present implementation is the step of finding the numerical solution of the Pfaffian system within \textsc{Mathematica}. A native \texttt{C++} integrator based on the Bulirsch–Stoer algorithm, interfaced with \texttt{HyperPrecision}, would substantially reduce the runtime.

    \item High-precision numerical evaluations would, in turn, open the possibility of reconstructing analytic expressions for MHFs from numerical data using integer-relation algorithms such as PSLQ~\cite{Ferguson:1992,Ferguson:1999} or lattice-reduction algorithms based on LLL~\cite{Lenstra:1982,Bailey:1999nv}, following the same general strategy as recent reconstructions of Feynman integrals from high-precision samples~\cite{Chachamis:2008fx,Lee:2010cga,Barrera:2025uin}.

    \item Another limitation of the present version concerns branch selection during transport. When the straight-line contour crosses a real singularity of the Pfaffian system, the option \texttt{IDelta} specifies the side on which the singularity is avoided. In Feynman integral calculations, the physically correct side is determined by the causal $+i0$ prescription of the underlying integral. This prescription propagates through the change of variables and fixes the signs of the imaginary parts of the hypergeometric arguments, as illustrated for the one-loop bubble in Section~5.4.1. The present version does not determine this assignment automatically, so the user must set \texttt{IDelta} manually. A natural way to address this limitation is to extend the package to support complex target points. Assigning the relevant invariant its physical infinitesimal imaginary part moves the target away from the real singular locus, ensuring that the transport contour no longer crosses a real singularity. The complex deformation then selects the correct branch automatically, without requiring a manual choice of \texttt{IDelta}. We therefore view the automatic determination of \texttt{IDelta} and support for complex target points as two aspects of the same future development.
\end{enumerate}

Given the widespread appearance of MHFs in mathematical and high-energy physics, and the range of examples covered by the present implementation, we expect \texttt{HyperPrecision} to be a useful tool for the broader scientific community.

\paragraph{Acknowledgments:}
S. Bera thanks Martijn Hidding for discussions related to this work at the workshop
``MathemAmplitudes 2023: QFT at the Computational Frontier''
and for his contributions during the initial stages of this work.
S. Bera also thanks Arpita Mitra and Debangshu Mukherjee for fruitful discussions on cosmological correlators, and Anna-Laura Sattelberger for related discussions. We thank the authors of \cite{Aoki:2024uyi} for helpful email correspondence.
We also thank the authors of~\cite{Haug:2025sre} for providing a Wolfram notebook that cross-checks our numerical evaluation of the three-denominator massless angular integral against their analytic result.
We are grateful to Taushif Ahmed, B. Ananthanarayan, Sudeepan Datta, Lance Dixon, Samuel Friot, Xin Guan, Syed Mehedi Hasan, Heribertus Bayu Hartanto, M.S.A. Alam Khan, Tanay Pathak, and Ratan Sarkar for carefully reading the manuscript and providing useful comments.
S. Banik gratefully acknowledges the financial support of the Swiss National Science Foundation Project No.\ P500-2\_235325. S. Bera has been supported by an appointment to the JRG Program at the APCTP through the Science and Technology Promotion Fund and Lottery Fund of the Korean Government and by the Korean Local Governments~--~Gyeongsangbuk-do Province and Pohang City.

\appendix

\section{Additional Function Reference}\label{app:functions}

This appendix collects the calling conventions, arguments, return values, and (where applicable) options for all the commands of \texttt{HyperPrecision.wl}, other than \cmd{HypExpand[]} and \cmd{HypFunctionExpand[]}, which are documented in the main text. One can also obtain the usage of the commands by running \texttt{?Function} in \textsc{Mathematica}.

\hspace{1cm} Several of the commands below share the same first three arguments, which we define once here and refer to throughout the appendix:
\begin{itemize}
    \item \texttt{HypSeries}~--~the symbolic hypergeometric series given in terms of Pochhammer symbols;
    \item \texttt{SumVar}~--~the list of summation indices appearing in \texttt{HypSeries};
    \item \texttt{Vars}~--~the list of kinematic variables.
\end{itemize}

\subsection{Generating Annihilating Partial Differential Equations:
\texorpdfstring{\\}{ } 
\cmd{PDEGenerator[]}}\label{app:PDEGenerator}

\begin{lstlisting}[language=Mathematica]
PDEGenerator[HypSeries, SumVar, Vars, FuncSymbol]
\end{lstlisting}

\paragraph{Description:} Derives the list of partial differential equations satisfied by \texttt{HypSeries} in the variables \texttt{Vars}, derived from the shift ratios of the series coefficients in the summation indices \texttt{SumVar}.

\paragraph{Arguments:} The first three arguments are as defined at the start of the appendix. The optional fourth argument \texttt{FuncSymbol} (default \texttt{F}) sets the function symbol that appears in the output partial differential equations.

\paragraph{Return value:} The list of partial differential equations in the unknown function \texttt{FuncSymbol[Vars]}.

\subsection{Deriving Order of Partial Differential Equations: \texorpdfstring{\\}{ } \cmd{FindHypergeometricOrder[]}}\label{app:FindHypergeometricOrder}

\begin{lstlisting}[language=Mathematica]
FindHypergeometricOrder[HypSeries, SumVar, Vars]
\end{lstlisting}

\paragraph{Description:} Derives the list of partial differential equation orders satisfied by \texttt{HypSeries}, one per summation index, derived from the polynomial degrees of the shift ratios of the series coefficients.

\paragraph{Return value:} A list of $k$ integers, where $k$ is the number of summation indices and the $j$-th entry gives the partial differential equation order in the direction of the $j$-th summation index.

\subsection{Holonomic Rank: \cmd{FindHolonomicRank[]}}\label{app:FindHolonomicRank}
\begin{lstlisting}[language=Mathematica]
FindHolonomicRank[HypSeries, SumVar, Vars]
\end{lstlisting}
\paragraph{Description:} Computes the holonomic rank of the Pfaffian system associated with \texttt{HypSeries}, together with a basis realizing this rank.

\paragraph{Return value:} The pair \texttt{\{N, Basis\}}, where \texttt{N} is the holonomic rank and \texttt{Basis} is a list of \texttt{N} independent derivatives of the unknown function \texttt{F}, sorted by total derivative order. The returned \texttt{Basis} can be passed directly to \cmd{FindPfaffianSystemBasis[]} as the user-supplied basis argument.

\subsection{Deriving Pfaffian: \cmd{FindPfaffianSystem[]}}\label{app:FindPfaffianSystem}

\begin{lstlisting}[language=Mathematica]
FindPfaffianSystem[HypSeries, SumVar, Vars]
\end{lstlisting}

\paragraph{Description:} Derives the Pfaffian system associated with \texttt{HypSeries}, automatically constructing a basis of derivatives from the holonomic order of the system.

\paragraph{Return value:} The 4-tuple \texttt{\{HypSeries, SumVar, Pfaffian, Basis\}}, where \texttt{Pfaffian} is the Association \texttt{<|Var $\to$ Matrix|>} of connection matrices and \texttt{Basis} is the list of basis elements. The 4-tuple is the input format expected by \cmd{TransportDE[]} (see Appendix~\ref{app:TransportDE}).

\subsection{Deriving Pfaffian for a Basis: \cmd{FindPfaffianSystemBasis[]}}\label{app:FindPfaffianSystemBasis}

\begin{lstlisting}[language=Mathematica]
FindPfaffianSystemBasis[HypSeries, SumVar, Vars, Basis]
\end{lstlisting}

\paragraph{Description:} Same as \cmd{FindPfaffianSystem[]} (Appendix~\ref{app:FindPfaffianSystem}), except that the basis of derivatives is supplied by the user as the fourth argument \texttt{Basis} rather than constructed automatically.

\paragraph{Arguments:} The first three arguments are as defined at the start of the appendix. \texttt{Basis} is the user-supplied list of basis elements, given as derivatives of the unknown function \texttt{F}, e.g.\ \texttt{\{F[x,y], Derivative[1,0][F][x,y], Derivative[0,1][F][x,y], Derivative[1,1][F][x,y]\}}.

\paragraph{Return value:} The 4-tuple \texttt{\{HypSeries, SumVar, Pfaffian, Basis\}}, in the input format expected by \cmd{TransportDE[]}.

\begingroup
\renewcommand{\cmd}[1]{\texttt{{#1}}}

\subsection{Pre-Restricted Pfaffian: \cmd{FindRestrictedPfaffianSystem[]}}
\label{app:FindRestrictedPfaffianSystem}

\begin{lstlisting}[
    language=Mathematica,
    literate={}
]
FindRestrictedPfaffianSystem[HypSeries, SumVar, TargetRules, StartRules]
\end{lstlisting}

\paragraph{Description:} Constructs the restricted matrix $M(\eta)$ in~\eeqref{eqn:etaODE} directly along a contour from the starting point to the target point, without deriving the full multivariate Pfaffian system. This approach is more efficient when only one target point is required, especially at high holonomic rank.

\paragraph{Arguments:} \texttt{HypSeries} and \texttt{SumVar} are defined at the start of the appendix. All Pochhammer parameters in \texttt{HypSeries} must be assigned numerical values. \texttt{TargetRules} specifies the target point at $\eta=0$, whereas \texttt{StartRules} specifies a regular starting point away from the singular curves. In the three-argument form, the starting point defaults to the origin. An optional final argument can be used to supply a basis in the $\theta$-operator form used by \cmd{FindPfaffianSystem[]}; otherwise, the basis is constructed automatically.

\paragraph{Return value:} The 8-tuple
\texttt{\{HypSeries, SumVar, M, Basis, ContourSub, TargetRules, StartRules, LinearContourQ\}},
where \texttt{M} is the restricted matrix, \texttt{Basis} is the basis of the system, \texttt{ContourSub} parametrises the kinematic variables in terms of $\eta$, and \texttt{LinearContourQ} records whether the contour is linear. The contour parameter is the global symbol \texttt{eta}, which is shared with \texttt{DESolver} and must remain unassigned by the user. The result can be passed directly to \cmd{TransportDE[]}.

\paragraph{Options:} The option \cmd{LinearContour}, whose default value is \texttt{False}, selects the contour parametrisation
\[
x_i=x_i^{\mathrm{target}}
 +(x_i^{\mathrm{start}}-x_i^{\mathrm{target}})\eta
\]
when set to \texttt{True}, and
\[
x_i=x_i^{\mathrm{start}}
 +\frac{x_i^{\mathrm{target}}-x_i^{\mathrm{start}}}{1+\eta}
\]
when set to \texttt{False}. In the former case, the starting point corresponds to $\eta=1$; in the latter, it corresponds to $\eta\to\infty$. In both cases, the target point is at $\eta=0$.

\endgroup

\subsection{Frobenius Integrability Check: \cmd{CheckIntegrability[]}}\label{app:CheckIntegrability}

\begin{lstlisting}[language=Mathematica]
CheckIntegrability[PfaffianIN, Vars]
\end{lstlisting}

\paragraph{Description:} Checks the Frobenius integrability conditions
\begin{align}
\partial_j M_i - \partial_i M_j + [M_i, M_j] \;=\; 0
\end{align}
for every pair of variables in \texttt{Vars}, where the connection matrices $M_i$ are taken from the Pfaffian system \texttt{PfaffianIN}. The check is performed numerically at a randomly chosen rational point.

\paragraph{Arguments:} \texttt{PfaffianIN} is the 4-tuple returned by \cmd{FindPfaffianSystem[]} or \\ \cmd{FindPfaffianSystemBasis[]}.

\paragraph{Return value:} Prints \texttt{PASSED} or \texttt{FAILED} for each pair, and returns \texttt{True} if all pairs pass, \texttt{False} otherwise.

\subsection{Numerical Transport: \cmd{TransportDE[]}}\label{app:TransportDE}

\begin{lstlisting}[language=Mathematica, escapeinside={(*@}{@*)}]
TransportDE[PfaffianIN, ParSub, VarSub, {Eps, Ord}, NPrec,
            (* options *)
            IDelta       -> -I,
            VerboseMode  -> False,
            ParallelRun  -> True,
            SummedBoundary -> False,
            WPrecision -> Automatic]
\end{lstlisting}

\paragraph{Description:} Computes the Laurent expansion in \texttt{Eps} up to order \texttt{Ord} of the function defined by the Pfaffian system \texttt{PfaffianIN}, evaluated at the target point specified by \texttt{VarSub} to \texttt{NPrec} correct digits.

\paragraph{Arguments:} \texttt{PfaffianIN} is the value returned by \cmd{FindPfaffianSystem[]} or \\
\cmd{FindPfaffianSystemBasis[]}, or by \cmd{FindRestrictedPfaffianSystem[]}. \texttt{ParSub} is the list of substitution rules for the Pochhammer parameters. \texttt{VarSub} is the list of rules \texttt{Var -> TargetPoint} specifying the kinematic variables and the evaluation point. \texttt{Eps} is the symbol of the Laurent parameter, \texttt{Ord} is the highest order of the Laurent expansion to be computed, and \texttt{NPrec} sets the target number of correct digits in the final numerical result.

\paragraph{Return value:} The list of Laurent coefficients from the lowest non-vanishing order to \texttt{Ord}.

\paragraph{Options:} \cmd{TransportDE[]} accepts the options \cmd{IDelta}, \cmd{VerboseMode}, \cmd{ParallelRun}, \cmd{SummedBoundary}, and \cmd{WPrecision}, as described in the main text (Section~\ref{subsec:hypexpand}). Unlike \cmd{HypExpand[]} and \cmd{HypFunctionExpand[]}, \cmd{TransportDE[]} does not accept \cmd{PreRestriction}, because the input Pfaffian system is supplied directly.

\section{Seed Integrals for Cosmological Correlator}\label{app:seed_ints}

In this appendix, we collect the explicit expressions for the double massive exchange seed integrals introduced in Eqs.~(3.42)–(3.44) of \cite{Aoki:2024uyi}, which constitute the basic building blocks entering the cosmological correlators discussed in Section~\ref{sec:correlators}. Since the analytic structure of these integrals is rather involved, we first summarise the notation and conventions adopted throughout this appendix.

\hspace{1cm}  The seed integrals $\mathcal{I}^{p_1 p_2 p_3}_{\pm\mp\pm,\alpha\beta}$, $\mathcal{I}^{p_1 p_2 p_3}_{\pm\pm\mp,\alpha\beta}$, and $\mathcal{I}^{p_1 p_2 p_3}_{\pm\pm\pm,\alpha\beta}$
are classified according to their Schwinger--Keldysh (SK) index structure. The three $\pm$ signs appearing in the subscript denote the SK vertex indices, each taking values in $\{+,-\}$. The labels $\alpha$ and $\beta$ distinguish the two exchanged massive scalar fields $\sigma_\alpha$ and $\sigma_\beta$ and their masses, $m_\alpha$ and $m_\beta$, are encoded through the dimensionless parameters

\[\mu_\alpha = \sqrt{\frac{m_\alpha^2}{H^2} - \frac94},
\qquad
\mu_\beta = \sqrt{\frac{m_\beta^2}{H^2} - \frac94},
\]

where $H$ denotes the Hubble parameter.

\hspace{1cm} The integer indices $p_1$, $p_2$, and $p_3$ characterise the different types of $\phi$--$\sigma$ interaction vertices appearing in the seed integrals, with $\phi$ representing the massless inflaton fluctuation. For notational convenience, we employ the shorthand definitions
\[
p_{ij} \equiv p_i + p_j,
\qquad
p_{123} \equiv p_1 + p_2 + p_3.
\]

The kinematic dependence of the seed integrals is encoded in two equivalent sets of momentum ratios. The first consists of
\[
u \equiv \frac{k_1}{k_{24}},
\qquad
v \equiv \frac{k_3}{k_{24}},
\]
where $k_i \equiv |\mathbf{k}_i|$ denote the magnitudes of the external momenta and $k_{24} \equiv k_2 + k_4$. The second set is given by the rescaled variables
\[
r_1 \equiv \frac{2k_1}{k_{1234}},
\qquad
r_3 \equiv \frac{2k_3}{k_{1234}},
\]
with
\[
k_{1234} \equiv k_1 + k_2 + k_3 + k_4.
\]
These two parameterisations are related through
\[
r_1 = \frac{2u}{1+u+v},
\qquad
r_3 = \frac{2v}{1+u+v}.
\]

With these conventions established, we now present the explicit forms of the seed integrals below. The derivation of the seed integrals can be found in \cite{Aoki:2024uyi}.

\begin{small}
\begin{align}
&\mathcal{I}^{p_1 p_2 p_3}_{\pm\mp\pm,\alpha\beta}\\
&= -\,e^{\mp i\frac{\pi}{2}(p_{13}-p_2)}\,
\widetilde{\Gamma}(p_1, p_2, p_3, \mu_\alpha, \mu_\beta)
\sum_{a,b=\pm}
\operatorname{csch}(\pi a\mu_\alpha)\,
\operatorname{csch}(\pi b\mu_\beta)\,
u^{-\frac{5}{2}-p_1-ia\mu_\alpha}\,
v^{-\frac{5}{2}-p_3-ib\mu_\beta}
\nonumber\\[6pt]
&\quad\times
\frac{
  \Gamma\!\left(\dfrac{4+p_2-i(a\mu_\alpha+b\mu_\beta)}{2}\,, \dfrac{5+p_2-i(a\mu_\alpha+b\mu_\beta)}{2}\right)
}{
  \Gamma\!\left(1-ia\mu_\alpha\,,1-ib\mu_\beta\right)
}
\nonumber\\[6pt]
&\quad\times
F_4\!\left(
  \frac{4+p_2-i(a\mu_\alpha+b\mu_\beta)}{2},\;
  \frac{5+p_2-i(a\mu_\alpha+b\mu_\beta)}{2};\;
  1-ia\mu_\alpha,\;
  1-ib\mu_\beta;\;
  u^2,\, v^2
\right).
\label{eq:3.42}
\end{align}

\end{small}

and, 

\begin{small}

\begin{align}
&\mathcal{I}^{p_1 p_2 p_3}_{\pm\pm\mp,\alpha\beta}\\
&= \mp i\,e^{\mp i\frac{\pi}{2}(p_{12}-p_3)}\,
\widetilde{\Gamma}(p_1,p_2,p_3,\mu_\alpha,\mu_\beta)
\sum_{a,b=\pm}
\operatorname{csch}(\pi a\mu_\alpha)\,
\operatorname{csch}(\pi b\mu_\beta)\,
e^{\mp\pi a\mu_\alpha}\,
u^{-\frac{5}{2}-p_1-ia\mu_\alpha}\,
v^{-\frac{5}{2}-p_3-ib\mu_\beta}
\nonumber\\[6pt]
&\quad\times
\frac{
  \Gamma\!\left(\dfrac{4+p_2-i(a\mu_\alpha+b\mu_\beta)}{2}\,,\dfrac{5+p_2-i(a\mu_\alpha+b\mu_\beta)}{2}\right)
}{
  \Gamma(1-ia\mu_\alpha\,, 1-ib\mu_\beta)
}
\nonumber\\[6pt]
&\quad\times
F_4\!\left(
  \frac{4+p_2-i(a\mu_\alpha+b\mu_\beta)}{2},\;
  \frac{5+p_2-i(a\mu_\alpha+b\mu_\beta)}{2};\;
  1-ia\mu_\alpha,\;1-ib\mu_\beta;\;
  u^2,\,v^2
\right)
\nonumber\\[12pt]
&\quad+\;
\frac{i}{\pi^{1/2}}\,
\frac{r_3^{p_{12}+\frac{13}{2}}}{v^{p_{123}+9}}\,
\frac{e^{\mp i\frac{\pi}{2}(p_{12}-p_3)}}{2^{6+2p_2}}\,
\widetilde{\Gamma}(p_1,p_2,p_3,\mu_\alpha,\mu_\beta)\,
\frac{
  \Gamma(3+p_1)
}{
  \Gamma\!\left(\dfrac{7}{2}+p_1-i\mu_\alpha\,,
  \dfrac{7}{2}+p_1+i\mu_\alpha\right)
}
\nonumber\\[6pt]
&\quad\times
\sum_{a=\pm}
\operatorname{csch}(\pi a\mu_\beta)\,
r_3^{\,ia\mu_\beta}
\sum_{m=0}^{\infty}\;\sum_{n=0}^{\infty}
\frac{ (1)_m
  (3+p_1)_m
  \left(\dfrac{1}{2}+ia\mu_\beta\right)_{\!n}
  \left(\dfrac{13}{2}+m+p_{12}+ia\mu_\beta\right)_{\!n}
}{
  \!\left(\dfrac{7}{2}+p_1-i\mu_\alpha\right)_{\!m}
  \left(\dfrac{7}{2}+p_1+i\mu_\alpha\right)_{\!m}
\left(1+2ia\mu_\beta\right)_{\!n}\, m! n!
}
\,
r_1^m\,r_3^n.
\label{eq:3.43_double}
\end{align}

\end{small}

and,
\begin{small}

\begin{align}
&\mathcal{I}^{p_1 p_2 p_3}_{\pm\pm\pm,\alpha\beta}\\
&= e^{\mp i\frac{\pi}{2}p_{123}}\,
\widetilde{\Gamma}(p_1,p_2,p_3,\mu_\alpha,\mu_\beta)
\sum_{a,b=\pm}
\operatorname{csch}(\pi a\mu_\alpha)\,
\operatorname{csch}(\pi b\mu_\beta)\,
e^{\mp\pi(a\mu_\alpha+b\mu_\beta)}\,
u^{-\frac{5}{2}-p_1-ia\mu_\alpha}\,
v^{-\frac{5}{2}-p_3-ib\mu_\beta}
\nonumber\\[6pt]
&\quad\times
\frac{
  \Gamma\!\left(\dfrac{4+p_2-i(a\mu_\alpha+b\mu_\beta)}{2}\,,\dfrac{5+p_2-i(a\mu_\alpha+b\mu_\beta)}{2}\right)
}{
  \Gamma(1-ia\mu_\alpha\,,1-ib\mu_\beta)
}
\nonumber\\[6pt]
&\quad\times
F_4\!\left(
  \tfrac{4+p_2-i(a\mu_\alpha+b\mu_\beta)}{2},\;
  \tfrac{5+p_2-i(a\mu_\alpha+b\mu_\beta)}{2};\;
  1-ia\mu_\alpha,\;1-ib\mu_\beta;\;
  u^2,\,v^2
\right)
\nonumber\\[16pt]
&\quad+\;\Bigg\{
\phantom{+\Bigg\{}
\frac{r_3^{\,p_{123}+9}}{v^{p_{123}+9}}
\cdot
\frac{\Gamma(9+p_{123})\,e^{\mp i\frac{\pi}{2}p_{123}}}
     {2^{10+p_{123}}
      \bigl[\mu_\alpha^2+\bigl(p_1+\tfrac{5}{2}\bigr)^2\bigr]
      \bigl[\mu_\beta^2 +\bigl(p_3+\tfrac{5}{2}\bigr)^2\bigr]}
\nonumber\\[6pt]
&\qquad\times
\sum_{m=0}^{\infty}\sum_{n=0}^{\infty}
\frac{ (1)_m (1)_n
  (3+p_1)_m\;
  (3+p_3)_n\;
  (9+p_{123})_{m+n}
}{
  \bigl(\tfrac{7}{2}+p_1-i\mu_\alpha\bigr)_{\!m}\,
  \bigl(\tfrac{7}{2}+p_1+i\mu_\alpha\bigr)_{\!m}\,
  \bigl(\tfrac{7}{2}+p_3-i\mu_\beta\bigr)_{\!n}\,
  \bigl(\tfrac{7}{2}+p_3+i\mu_\beta\bigr)_{\!n }
} \frac{\,r_1^m\,r_3^n}{m!n!}
\nonumber\\[16pt]
&\quad\pm\;
\sum_{a=\pm}
\bigl(\coth(\pi a\mu_\beta)\mp 1\bigr)\,
\frac{r_3^{\,p_{12}+\frac{13}{2}-ia\mu_\beta}}{v^{p_{123}+9}}
\cdot
\frac{e^{\mp i\frac{\pi}{2}p_{123}}}
     {2^{10+p_{123}}
      \bigl[\mu_\alpha^2+\bigl(p_1+\tfrac{5}{2}\bigr)^2\bigr]}
\nonumber\\[6pt]
&\qquad\times
\frac{
  \Gamma\!\bigl(\tfrac{5}{2}+p_3-i\mu_\beta\,, \tfrac{5}{2}+p_3+i\mu_\beta\,,\tfrac{1}{2}-ia\mu_\beta\,,\tfrac{13}{2}+p_{12}-ia\mu_\beta\bigr)
}{
  \Gamma(3+p_3\,,1-2ia\mu_\beta)
}
\nonumber\\[6pt]
&\qquad\times
\sum_{m=0}^{\infty}\sum_{n=0}^{\infty}
\frac{ (1)_m
  (3+p_1)_m\;
  \bigl(\tfrac{1}{2}-ia\mu_\beta\bigr)_{\!n}\;
  \bigl(\tfrac{13}{2}+p_{12}-ia\mu_\beta\bigr)_{\!m+n}
}{
  \bigl(\tfrac{7}{2}+p_1-i\mu_\alpha\bigr)_{\!m}\,
  \bigl(\tfrac{7}{2}+p_1+i\mu_\alpha\bigr)_{\!m}\,
  (1-2ia\mu_\beta)_{\!n}\cdot 
} \frac{\,r_1^m\,r_3^n}{m!n!}
\nonumber\\[12pt]
&\quad+\;
\left(
  \mu_\alpha\leftrightarrow\mu_\beta,\quad
  p_1\leftrightarrow p_3,\quad
  r_1\leftrightarrow r_3
\right)\Bigg\}.
\label{eq:3.44_double}
\end{align}
    
\end{small}

where,

\begin{small}
    \begin{equation}
\widetilde{\Gamma}(p_1, p_2, p_3, \mu_\alpha, \mu_\beta)
\equiv
\frac{\pi^{1/2}}{2^{4+p_{13}-p_2}}
\frac{
  \Gamma\!\left(\dfrac{5}{2}+p_1-i\mu_\alpha,
  \dfrac{5}{2}+p_1+i\mu_\alpha,\dfrac{5}{2}+p_3-i\mu_\beta,\dfrac{5}{2}+p_3+i\mu_\beta\right)
}{
  \Gamma(3+p_1,3+p_3)
}.
\label{eq:3.41}
\end{equation}
\end{small}

The expression for $\mathcal{I}^{p_1 p_2 p_3}_{\pm\mp\mp,\alpha\beta}$ can be obtained by 
replacing $u \leftrightarrow v$ ($r_1 \leftrightarrow r_3$), $p_1 \leftrightarrow p_3$, 
and $\alpha \leftrightarrow \beta$ in $\mathcal{I}^{p_1 p_2 p_3}_{\mp\mp\pm,\alpha\beta}$. In the above expressions, we have used the notation:
\begin{align*}
    \Gamma(a_1,a_2,\dots, a_n) = \prod_{i=1}^n \Gamma(a_i)
\end{align*}

\section{Predefined Hypergeometric Functions}
\label{app:predefined}

This appendix lists the named MHFs recognised by \cmd{HypFunctionExpand[]}, together with their series definitions and calling conventions. We recall that \texttt{HyperPrecision} is not limited to these predefined functions, since users can also evaluate their own Horn-type series, written explicitly in terms of Pochhammer symbols, using \cmd{HypExpand[]}.

\subsection{Gauss and Generalised Hypergeometric Functions}
\label{app:gauss}

The Gauss hypergeometric functions $_2F_1$ and $_pF_q$ are built-in commands of \textsc{Mathematica}, and are defined by

\begin{align}
{}_2F_1(a,b;c;x)
&=
\sum_{m\geq 0}
\frac{(a)_m(b)_m}{(c)_m\,m!}\,x^m,
\\
{}_pF_q(a_1,\ldots,a_p;b_1,\ldots,b_q;x)
&=
\sum_{m\geq 0}
\frac{\prod_{i=1}^{p}(a_i)_m}
     {\prod_{j=1}^{q}(b_j)_m\,m!}\,x^m .
\end{align}
The corresponding calling conventions are
\begin{lstlisting}[language=Mathematica]
Hypergeometric2F1[a, b, c, x]
HypergeometricPFQ[{a1, ..., ap}, {b1, ..., bq}, x]
\end{lstlisting}
We note that for $_pF_q$, our package can only evaluate for the case $q=p-1$ for any $p\geq 2$. 

\subsection{Appell Functions}
\label{app:appell}

The Appell functions $F_1$, $F_2$, $F_3$ and $F_4$ are also built-in commands of \textsc{Mathematica}, and are defined by
\begin{align}
F_1(a;b_1,b_2;c;x,y)
&=
\sum_{m,n\geq 0}
\frac{(a)_{m+n}(b_1)_m(b_2)_n}
     {(c)_{m+n}\,m!\,n!}\,
x^m y^n,
\\
F_2(a;b_1,b_2;c_1,c_2;x,y)
&=
\sum_{m,n\geq 0}
\frac{(a)_{m+n}(b_1)_m(b_2)_n}
     {(c_1)_m(c_2)_n\,m!\,n!}\,
x^m y^n,
\\
F_3(a_1,a_2;b_1,b_2;c;x,y)
&=
\sum_{m,n\geq 0}
\frac{(a_1)_m(a_2)_n(b_1)_m(b_2)_n}
     {(c)_{m+n}\,m!\,n!}\,
x^m y^n,
\\
F_4(a,b;c_1,c_2;x,y)
&=
\sum_{m,n\geq 0}
\frac{(a)_{m+n}(b)_{m+n}}
     {(c_1)_m(c_2)_n\,m!\,n!}\,
x^m y^n .
\end{align}
The corresponding calling conventions are
\begin{lstlisting}[language=Mathematica]
AppellF1[a, b1, b2, c, x, y]
AppellF2[a, b1, b2, c1, c2, x, y]
AppellF3[a1, a2, b1, b2, c, x, y]
AppellF4[a, b, c1, c2, x, y]
\end{lstlisting}

\subsection{Horn \texorpdfstring{$G$}{G}-Series}
\label{app:hornG}

The Horn $G$-series $G_1$, $G_2$ and $G_3$ are provided by \texttt{HyperPrecision} and are defined by
\begin{align}
G_1(a,b,c;x,y)
&=
\sum_{m,n\geq 0}
\frac{(a)_{m+n}(b)_{n-m}(c)_{m-n}}
     {m!\,n!}\,
x^m y^n,
\\
G_2(a,b,c,d;x,y)
&=
\sum_{m,n\geq 0}
\frac{(a)_m(b)_n(c)_{n-m}(d)_{m-n}}
     {m!\,n!}\,
x^m y^n,
\\
G_3(a,b;x,y)
&=
\sum_{m,n\geq 0}
\frac{(a)_{2n-m}(b)_{2m-n}}
     {m!\,n!}\,
x^m y^n .
\end{align}
The corresponding calling conventions are
\begin{lstlisting}[language=Mathematica]
HornG1[a, b, c, x, y]
HornG2[a, b, c, d, x, y]
HornG3[a, b, x, y]
\end{lstlisting}

\subsection{Horn \texorpdfstring{$H$}{H}-Series}
\label{app:hornH}

The Horn $H$-series $H_1,\ldots,H_7$ are also provided by \texttt{HyperPrecision} and are defined by
\begin{align}
H_1(a,b,c;d;x,y)
&=
\sum_{m,n\geq 0}
\frac{(a)_{m-n}(b)_{m+n}(c)_n}
     {(d)_m\,m!\,n!}\,
x^m y^n,
\\
H_2(a,b,c,d;e;x,y)
&=
\sum_{m,n\geq 0}
\frac{(a)_{m-n}(b)_m(c)_n(d)_n}
     {(e)_m\,m!\,n!}\,
x^m y^n,
\\
H_3(a,b;c;x,y)
&=
\sum_{m,n\geq 0}
\frac{(a)_{2m+n}(b)_n}
     {(c)_{m+n}\,m!\,n!}\,
x^m y^n,
\\
H_4(a,b;c,d;x,y)
&=
\sum_{m,n\geq 0}
\frac{(a)_{2m+n}(b)_n}
     {(c)_m(d)_n\,m!\,n!}\,
x^m y^n,
\\
H_5(a,b;c;x,y)
&=
\sum_{m,n\geq 0}
\frac{(a)_{2m+n}(b)_{n-m}}
     {(c)_n\,m!\,n!}\,
x^m y^n,
\\
H_6(a,b,c;x,y)
&=
\sum_{m,n\geq 0}
\frac{(a)_{2m-n}(b)_{n-m}(c)_n}
     {m!\,n!}\,
x^m y^n,
\\
H_7(a,b,c;d;x,y)
&=
\sum_{m,n\geq 0}
\frac{(a)_{2m-n}(b)_n(c)_n}
     {(d)_m\,m!\,n!}\,
x^m y^n .
\end{align}
The corresponding calling conventions are
\begin{lstlisting}[language=Mathematica]
HornH1[a, b, c, d, x, y]
HornH2[a, b, c, d, e, x, y]
HornH3[a, b, c, x, y]
HornH4[a, b, c, d, x, y]
HornH5[a, b, c, x, y]
HornH6[a, b, c, x, y]
HornH7[a, b, c, d, x, y]
\end{lstlisting}

\subsection{Lauricella Functions}
\label{app:lauricella}

The Lauricella functions $F_A^{(n)}$, $F_B^{(n)}$, $F_C^{(n)}$ and $F_D^{(n)}$ in $n$ variables are also provided by \texttt{HyperPrecision} and are defined by
\begin{align}
F_A^{(n)}(a;b_1,\ldots,b_n;c_1,\ldots,c_n;x_1,\ldots,x_n)
&=
\sum_{m_1,\ldots,m_n\geq 0}
(a)_{m_1+\cdots+m_n}
\prod_{i=1}^{n}
\frac{(b_i)_{m_i}}{(c_i)_{m_i}\,m_i!}
x_i^{m_i},
\\
F_B^{(n)}(a_1,\ldots,a_n;b_1,\ldots,b_n;c;x_1,\ldots,x_n)
&=
\sum_{m_1,\ldots,m_n\geq 0}
\frac{1}{(c)_{m_1+\cdots+m_n}}
\prod_{i=1}^{n}
\frac{(a_i)_{m_i}(b_i)_{m_i}}{m_i!}
x_i^{m_i},
\\
F_C^{(n)}(a,b;c_1,\ldots,c_n;x_1,\ldots,x_n)
&=
\sum_{m_1,\ldots,m_n\geq 0}
(a)_{m_1+\cdots+m_n}
(b)_{m_1+\cdots+m_n}
\prod_{i=1}^{n}
\frac{x_i^{m_i}}{(c_i)_{m_i}\,m_i!},
\label{eqn:LauricellaFC}\\
F_D^{(n)}(a;b_1,\ldots,b_n;c;x_1,\ldots,x_n)
&=
\sum_{m_1,\ldots,m_n\geq 0}
\frac{(a)_{m_1+\cdots+m_n}}
     {(c)_{m_1+\cdots+m_n}}
\prod_{i=1}^{n}
\frac{(b_i)_{m_i}}{m_i!}
x_i^{m_i}.
\end{align}
The corresponding calling conventions are

\begin{lstlisting}[language=Mathematica]
LauricellaFA[{a, b1, ..., bn}, {c1, ..., cn}, {x1, ..., xn}]
LauricellaFB[{a1, ..., an, b1, ..., bn}, {c}, {x1, ..., xn}]
LauricellaFC[{a, b}, {c1, ..., cn}, {x1, ..., xn}]
LauricellaFD[{a, b1, ..., bn}, {c}, {x1, ..., xn}]
\end{lstlisting}

We note that for $n=2$, the Lauricella functions $F_A^{(2)}$, $F_B^{(2)}$, $F_C^{(2)}$ and $F_D^{(2)}$ correspond to the Appell functions $F_2$, $F_3$, $F_4$ and $F_1$, respectively.

\printbibliography

\end{document}